\documentclass[10pt,twocolumn,letterpaper]{article}

\usepackage[pagebackref,breaklinks,colorlinks]{hyperref}
\usepackage{translations}
\usepackage{romannum}
\usepackage{graphicx}	
\usepackage{amsmath}	
\usepackage{amssymb}	
\usepackage{multirow}
\usepackage{bm}
\usepackage{authblk}
\usepackage[capitalize]{cleveref}
\crefname{section}{Sec.}{Secs.}
\Crefname{section}{Section}{Sections}
\Crefname{table}{Table}{Tables}
\crefname{table}{Tab.}{Tabs.}
\usepackage{color}
\usepackage{caption}
\newcommand\kms{\ensuremath{\mbox{km}\,\mbox{s}^{-1}}}
\newcommand\Teff{\ensuremath{T_\mathrm{eff}}}
\newcommand\logg{\ensuremath{\log g}}
\newcommand\vsini{\ensuremath{v_{e}\sin i}}

\newcommand\met{\ensuremath{[M/H]}}

\begin{document}
\title{\LaTeX\ Deep Learning application for stellar parameters determination: \\
	I- Constraining the hyperparameters}	

\author{Marwan Gebran\\
	Department of Chemistry and Physics, Saint Mary’s College, Notre Dame, IN 46556, USA\\

	{\tt\small mgebran@saintmarys.edu}

\and
Kathleen Connick\\
Department of Chemistry and Physics, Saint Mary’s College, Notre Dame, IN 46556, USA\\

\and
Hikmat Farhat\\
Department of Computer Science, Notre Dame University-Louaize, PO Box 72, Zouk Mika\"{e}l, Lebanon\\

\and
Fr\'ed\'eric Paletou\\
Universit\'e de Toulouse, Observatoire Midi--Pyr\'en\'es, Irap, Cnrs, Cnes,  14 av. E. Belin, F--31400 Toulouse, France\\

\and
Ian Bentley\\
Department of Chemistry and Physics, Saint Mary’s College, Notre Dame, IN 46556, USA\\}

\maketitle

	\begin{abstract}
		Machine Learning is an efficient method for analyzing and interpreting the increasing amount of astronomical data that is available. In this study, we show, a pedagogical approach that should benefit anyone willing to experiment with Deep Learning techniques in the context of stellar parameters determination. Utilizing the Convolutional Neural Network architecture, we give a step by step overview of how to select the optimal parameters for deriving the most accurate values for the stellar parameters of stars: \Teff, \logg, \met, and \vsini. Synthetic spectra with random noise were used to constrain this method and to mimic the observations. We found that each stellar parameter requires a different combination of network hyperparameters and the maximum accuracy reached depends on this combination, as well as, the Signal to Noise ratio of the observations, and the architecture of the network. We also show that this technique can be applied to other spectral types in different wavelength ranges after the technique has been optimized.
	\end{abstract}

	\section{Introduction}
	Machine learning (ML) applications have been used extensively in astronomy over the last decade (\cite{2019arXiv190407248B}). This is mainly due to the large amount of data that are recovered from space and ground-based observatories. There is therefore a need to analyze this data in an automated way. Statistical approaches, dimensionality reduction, wavelet decomposition, ML, and Deep Learning (DL) are all examples of the attempts that were performed in order to derive more accurate stellar parameters such as the effective temperature (\Teff), surface gravity (\logg), projected equatorial rotational velocity (\vsini), and metallicity (\met) using stellar spectra in different wavelength ranges (\cite{Guiglion20,CARMENES1,2020AJ....160...45P,2020ApJ...891...23W,2020ApJS..246....9Z,2019AJ....158...93B,2019OAst...28...68K,Fabro,2018A&A...612A.111G,2017RAA....17...36L,Gebran,S4n,dms}). DL is a machine learning method based on deep Artificial Neural Networks (ANN) that does not usually require a specific statistical algorithm to predict a solution but it is rather learned by experience and thus require a very large dataset (\cite{cite-key}) for training in order to perform properly.

	An overview of the automated techniques used in stellar parameters determination can be found in \cite{2019OAst...28...68K}. We will mention some of the recent studies that involve ML/DL. The increase of the computational power and the large availability of predefined optimized ML packages (in e.g. Python, C++, and R) have allowed astronomers to shift from classical techniques to ML when using large data.
	One of the first trials to derive the stellar parameters using neural networks was done by \cite{1997PASP..109..932B}. This work demonstrated that networks can give accurate spectral type classifications across the spectral types range B2-M7.
	
	\cite{2016A&A...594A..68D} presented an ANN architecture that learns the function which can relate the stellar parameters to the input spectra. They obtained residuals in the derivation of the metallicity below 0.1 dex for stars with Gaia magnitude $\mathrm{G}_{rvs} < 12$ mag, which accounts for a number in the order of four million stars to be observed by the Radial Velocity Spectrograph of the Gaia satellite\footnote{The limited magnitude of the RVS is around 15.5 mag (\cite{2014EAS....67...69C}).}. \cite{2018A&A...619A..22R} used a ML algorithm to measure the mean longitudinal magnetic field in stars from polarized spectra of high resolution. They found a considerable improvement of the results, allowing to estimate
	the errors associated to the measurements of stellar magnetic fields at different noise levels. 
	
	\cite{2018MNRAS.476.1151P} developed, and applied a Convolutional Neural Network (CNN) architecture using multi-task learning to search for and characterize strong HI Ly$\alpha$ absorption in quasar spectra. \cite{Fabro} applied a deep neural network architecture to analyse both SDSS-III APOGEE DR13 and synthetic stellar spectra. This work demonstrated that the stellar parameters are determined with similar precision and accuracy as the APOGEE pipeline.
	
	\cite{2020MNRAS.491.2280S} introduced an automated approach for the classification of stellar spectra in the optical region using CNN. They also showed that deep learning methods with a larger number of layers allow the use of finer details in the spectrum which results in improved accuracy and better generalisation with respect to traditional ML techniques.
	
	\cite{2020ApJ...891...23W} introduced a deep-learning method, SPCANet, that derived \Teff\ and \logg\ and 13 chemical abundances for LAMOST Medium-Resolution Survey (MRS) data. These authors found abundance precision up to 0.19 dex for spectra with Signal-to-Noise ratio (SNR) down to $\sim$10. The results of SPCANet are consistent with those from other surveys such as APOGEE, GALAH and RAVE, and are also validated with the previous literature values including clusters and field stars. \cite{Guiglion20} derived the atmospheric parameters and abundances of different species for 420,165 RAVE spectra. They showed that CNN-based methods provide a powerful way to combine spectroscopic, photometric, and astrometric data without the need to apply any priors in the form of stellar evolutionary models. 
	
	More recently, \cite{2021arXiv210112550L} introduced a
	multi-layer CNN to forecast solar flare events probability occurrence of M and X classes. \cite{2021MNRAS.501.3951C}  introduced an AGN recognition method based on Deep Neural Network. \cite{2021arXiv210205809A} used machine learning ML methods to generate model SEDs and fit sparse observations of low-luminosity active galactic nuclei.  \cite{rhea202,rhea21} used CNNs and different ANN networks to estimate emission-line parameters and line ratios present in  different filters of  SITELLE spectrometer. \cite{Curran21} used DL combined with k-Nearest Neighbour and Decision Tree Regression algorithms to compare the accuracy of the predicted photometric redshifts of newly detected sources. \cite{2022NewA...9101693O} applied the \textit{ThetaRay} Artificial Intelligence algorithms to 10\,803 light curves of threshold crossing events and uncovered 39 new exoplanetary candidates targets. \cite{Bickley21} reached a classification accuracy of 88 percent while investigating the use of a CNN for automated merger classification. \cite{2021arXiv210309651G} used an assisted inversion techniques based on CNN for solar Stokes profile inversions. In the context of Classification of galactic morphologies, \cite{2021arXiv210309711G} used a ML generative adversarial networks to convert ground-based Subaru Telescope blurred images into quasi Hubble Space Telescope images. \cite{2021arXiv210607655G} presented StelNet, a Deep Neural Network trained on stellar evolutionary tracks that quickly
	and accurately predicts mass and age from absolute luminosity and effective temperature for stars of solar metallicity.
	
	In this manuscript, we present both a new method to derive stellar atmospheric parameters, and we also demonstrate the effect of each of the CNN parameters (such as the choice of the optimizers, loss function, activation function, $\ldots$) on the accuracy of the results. We will provide the procedure that can be followed in order to find the most appropriate configuration independently of the architecture of the CNN. This is intended as the first in a series of papers that will help the astronomical community to understand the effect on the accuracy of the prediction from most of the parameters and the architecture of the network. CNN parameters are numerous and to find the optimal ones is a very hard task. To do so, we trained the CNNs with different configurations of the parameters using purely synthetic spectra for the 3 steps of training, cross-validation (hereafter called validation), and testing. Using synthetic spectra, we have access to the true parameters during our tests. Noisy spectra are tested in order to mimic observations. 
	
	We have limited our work to a specific type of objects, A stars, because as mentioned previously the purpose is not to show how well we can derive the labeled stellar parameters but what is the effect of specific parameters on stellar spectra analysis. By applying our models to A stars, we use previous results (\cite{Gebran,2019OAst...28...68K}) as a reference for the expected accuracy of the derived stellar parameters. In the same way, the wavelength range and the resolving power are chosen to be representative of values used by most available instruments. Once the calibration of the hyperparameters was performed, we have tested our optimal network configurations on a set of FGK stars in Sec.~\ref{fgk}, using the wavelength range of \cite{S4n}. 
	
	The training, validation, and test data are explained in Sec.~\ref{LDB}. Section ~\ref{data} discusses the data preparation previous to training. The neural network construction and the parameters selection is explained in Sec.~\ref{DL}. Results are summarized in Sec.~\ref{results}. The application of the optimal networks to FGK stars is performed in Sec.~\ref{fgk}. 
	Discussion and conclusion are gathered in Sec.~\ref{disc}. 
	
	\section{Training spectra}
	\label{LDB}
	
	\begin{table}
		\centering
		\caption{Ranges of the parameters used for the calculation of the synthetic spectra TDB's.}
		\label{table_1}
		\begin{tabular}{|c|c|} 
			\hline
			Parameters & Range \\
			\hline
			\Teff\ (K) & [7\,000,11\,000]  \\
			\logg\ (dex)  & $[2.0, 5.0]$  \\
			\met   (dex)  & $[-1.5 , 1.5]$  \\ 
			\vsini\ (\kms) & $[0, 300] $  \\
			$\lambda/\Delta \lambda$ & 60\,000  \\
			\hline
		\end{tabular}
	\end{table}

	Our learning, or training databases (TDB) are constructed from synthetic spectra for stars having effective temperature between 7\,000 and 10\,000 K, and the wavelength range of 4450 \AA\ to 5000 \AA. This range was selected because it is in the visible domain and contains metallic and Balmer lines sensitive to all stellar parameters (\Teff, \logg, \met, \vsini), especially for the spectra types selected in this work. This region is also insensitive to microturbulent velocity which was adopted to be $\xi_t$=2 km/s based on the work of \cite{Gebran,micro}. Surface gravity, \logg, is selected to be in the range of 2.0--5.0 dex. Projected rotational velocity, \vsini, is calculated between 0--300 \kms. The Metallicity, \met, is in the range of -1.5 and +1.5 dex. Table ~\ref{table_1} displays the range of all stellar parameters. 
	These spectra are used for both the training and the validation phases. 
	Approximately 55\,000 noise free synthetic spectra were calculated using a random selection of the stellar parameters in the range of Tab.~\ref{table_1}. These spectra are used instead of the observations (Test data without noise). Gaussian SNR, ranging between 5 and 300, were added to these test spectra in order to check the accuracy of the technique on noisy data (Test data with noise).

	Details for the calculations of the synthetic spectra can be found in \cite{Gebran} or \cite{2019OAst...28...68K}. In summary, 1D plane--parallel model atmospheres were calculated using \texttt{ATLAS9} (\cite{Kurucz1992}). These models are in local thermodynamic equilibrium (LTE) and in hydrostatic and radiative equilibrium. We have used the new opacity distribution function in the calculations (\cite{castelli}) as well as a mixing length  parameter of 0.5 for 7000 K$\leq$\Teff$\leq$8500 K, and 1.25 for \Teff$\leq$7000 K (\cite{2004IAUS..224..131S}).
	
	We have used \cite{spectra} \texttt{SYNSPEC48} synthetic spectra code to calculate all normalized spectra. The adopted line lists are detailed in \cite{Gebran}. This list is mainly compiled using the data from Kurucz
	gfhyperall.dat\footnote{http://kurucz.harvard.edu}, VALD\footnote{http://www.astro.uu.se/$\sim$vald/php/vald.php}, and the
	NIST\footnote{http://physics.nist.gov} databases.
	
	Finally, the resolving power is simulated to \cal{R}=60\,000. This value falls in the range between low and high resolution spectrographs. The technique that will be shown in the next sections can be used for any resolution. 
	The construction and the size of the TDB will be discussed in Sec.~\ref{results}.
	The use of synthetic spectra in ML to constrain the stellar parameters has shown to suffer from the so-called synthetic gap (\cite{Fabro,CARMENES1}). This gap refers to the differences in feature distributions between synthetic and observed data. We have decided to limit our work to synthetic data for 2 reasons, first we would like to remove the hassle of the data preparation steps (data reduction, flux calibration, flux normalization, radial velocity correction $\ldots$), and second because our intention is to find the strategy and technique that should be adopted in ML for deriving stellar parameters.
	
	We are working on a future paper that deals with the architecture of the network as well as the choice of the kernel sizes and the number of neurons. Combining the best strategy to constrain the hyperparameters (this manuscript) as well as the most optimal architecture (future studies) will allow us to use a combination of synthetic and observational data in our training database. Having well known stellar parameters, these observational data will allow us to remove/minimize the synthetic gap and better constrain the stellar parameters.

	\section{Data preparation}
	\label{data}
\begin{figure}
		\centering
		\includegraphics[scale=0.6]{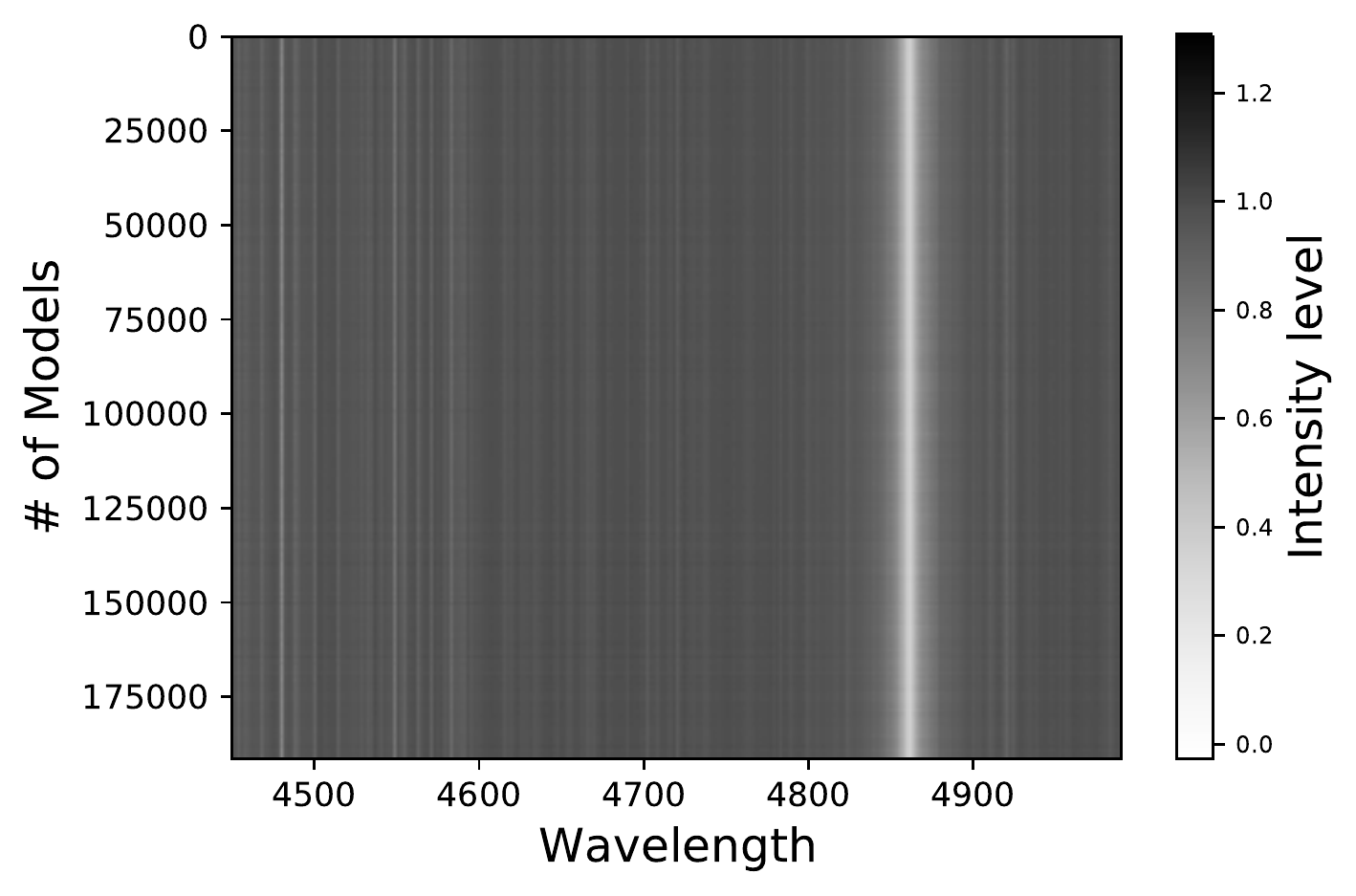}
		\caption{Color map representing the fluxes for a sample of the training database using data augmentation. Wavelengths are in \AA. \label{Colormap}}
	\end{figure}

	The TDB contains $N_{\mathrm{spectra}}$ spectra that span the wavelength range of 4450--5000 \AA. Having a wavelength step of 0.05 \AA, this results in $N_{\lambda}=$10800 flux points per spectrum. The TDB can then be represented by a matrix $\textbf{\textit{M}}$ of size $N_{\mathrm{spectra}} \times N_{\lambda}$.
	A color map of a subsample of $\textbf{\textit{M}}$ is displayed in Fig.\ref{Colormap}. Although the synthetic spectra are normalized, some wavelength points could have fluxes larger than unity. This is due to the noise that is incorporated during the so-called \emph{data augmentation} procedure which will be explained in Sec.~\ref{data_aug}.

	Training a CNN using the $\textbf{\textit{M}}$ matrix is time consuming, especially if one should use a larger wavelength range or a higher resolution. For that reason we have applied a dimensionality reduction technique i.e., Principal Component Analysis (PCA hereafter), in order to reduce the size of the training TDB as well as the size of the validation, test, and noisy synthetic data. Although this step is optional, we recommend its use whenever the data can be represented by a small number of coefficients. The PCA can reduce the size of each spectrum from $N_{\lambda}$ to $n_k$. The choice of $n_k$ depends on the many parameters, the size of the database, the wavelength range and the shape of the spectra lines. As a first step, we need to find the Principal Components, and to do so, we proceed as follows:\\
	
	The matrix $\textbf{\textit{M}}$ is averaged along the $N_{\mathrm{spectra}}$-axis and the result is stored in a vector $\bar{M}$. Then, we calculate the eigenvectors $\textbf{e}_k(\lambda)$ of the variance-covariance matrix $\textbf{\textit{C}}$  defined as 
	
	\begin{equation}
		\textbf{\textit{C}}=(\textbf{\textit{M}}-\bar{\textit{M}})^\mathrm{T}\cdot(\textbf{\textit{M}}-\bar{\textit{M}})\, 
	\end{equation}
	
	where the superscript "T" stands for the transpose operator. $\textbf{\textit{C}}$ has a dimension of $N_{\lambda}\times N_{\lambda}$. Sorting the eigenvectors of the variance-covariance matrix in decreasing magnitude will results in the "Principal Components".
	Each spectrum of $\textbf{\textit{M}}$ is then projected on these Principal Components in order to find its corresponding coefficient $p_{jk}$ defined as 
	\begin{equation}
		p_{jk}=(M_j-\bar{M})\cdot \textbf{\textit{e}}_k
	\end{equation}
	
	The choice of the number of coefficient is regulated by the reconstructed error as detailed in \cite{S4n}:
	
	\begin{equation}
		E(k_{max})=\left< \left( \dfrac{\vert\bar{M}+\Sigma_{k=1}^{k_{max}} p_{jk}\textbf{\textit{e}}_k - M_j\vert}{M_j}    \right)  \right>
	\end{equation}
	
\begin{center}
		\begin{figure}
			\includegraphics[scale=0.55]{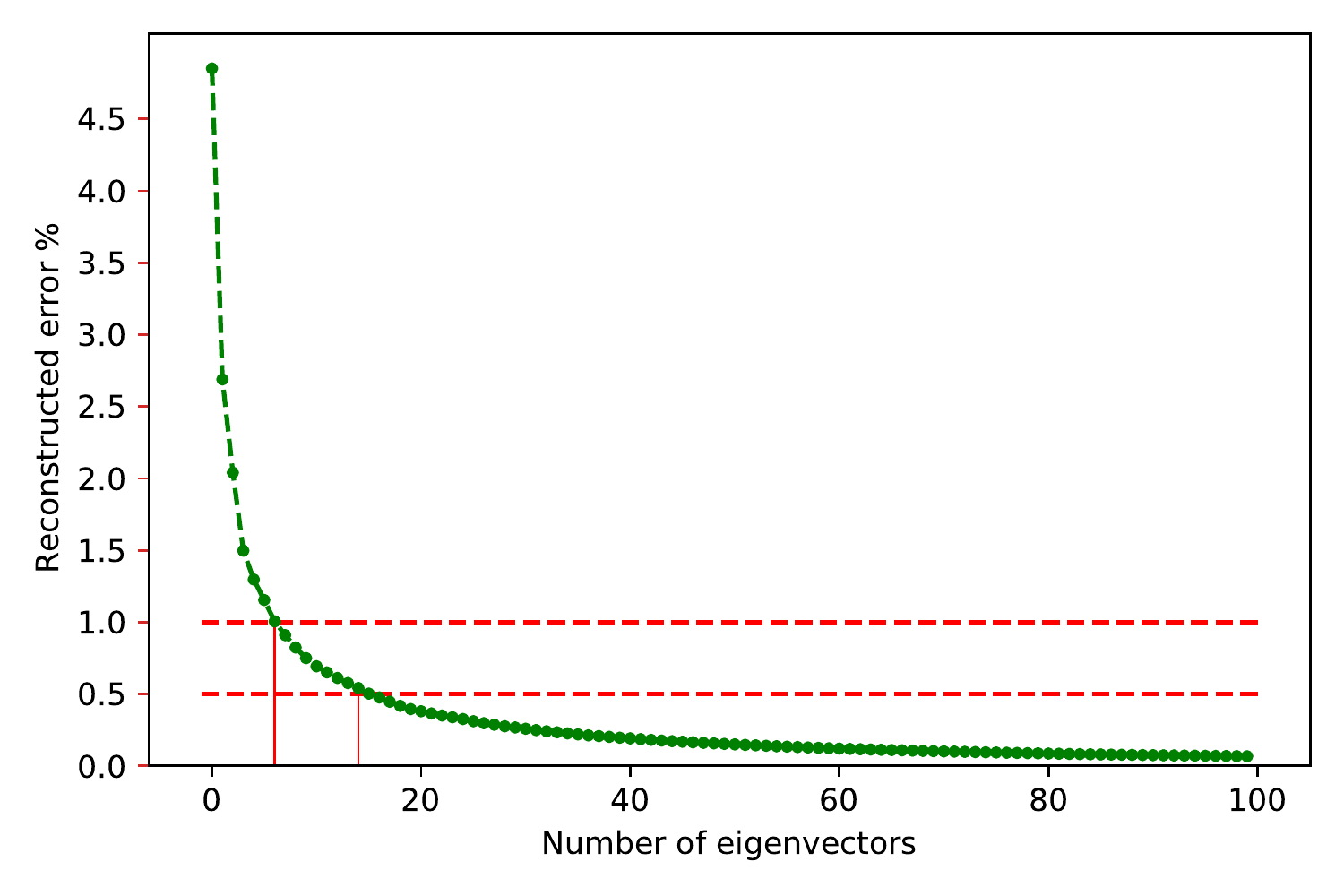}
			\caption{Mean reconstructed error as a function of the number of principal components used for the projection. The dashed lines represent the 1\% and 0.5\% error, respectively. For $n_k$> 17, the spectra can be reconstructed with more than 99.5\%  accuracy.}
			\label{PCA_error}
		\end{figure}
	\end{center}

	We have opted to a value for $n_k$ that reduces the mean reconstructed error to a value <0.5\%. As an example, using a database of 25\,000 spectra with stellar parameters ranging randomly between the values in Tab.~\ref{table_1} requires less than 7 coefficients to reach an accuracy <1\%, and a value of $n_k$=17 to reach a 0.5\% error as shown in Fig.~\ref{PCA_error}. This technique has shown its efficiency when applied to synthetic and/or real observational data with \Teff$>$ 4000 K (see \cite{Gebran, S4n,dms}, for more details).

	Applying the same procedure to all our TDB and taking the maximum value to be used for all, we have adopted a constant value for $n_k$=50. This value takes into account all the databases that will be dealt with in this work, especially that some will be data augmented as will be explained in Sec.~\ref{data_aug}. This means that instead of training a matrix having a dimension of $N_{\mathrm{spectra}} \times N_{\lambda}$, we are using one with dimension of $N_{\mathrm{spectra}} \times n_k$, with $n_k \ll N_{\lambda}$. In that case, our new data consist of a matrix containing the coefficients that are calculated by projecting the spectra on the $n_k$ eigenvectors.
	
	This projection procedure over the Principal Components is then applied to the validation, test, and noisy spectra datasets.

	\section{Deep Learning: Artificial Neural Network }
	\label{DL}
	
	This section begins with a brief description of supervised\footnote{Supervised learning refers to algorithms that calculate a predictive model using data points with known labels/outcomes.} learning. Given a data set $(X,Y)$, the goal is to find a function $f$ such that $f(X)$ is as "close" as possible to $Y$. For example, $Y$ could be the effective temperatures or surface gravity and $X$ the corresponding spectra. This "closeness" is typically measured by defining a \textit{loss function} $L(f(X),Y)$ that measures the difference between the predicted and actual values. Therefore the goal of the learning process is to find $f$ that minimizes $L$ for a given dataset $(X,Y)$. Ultimately, the success of any learning method is assessed by how well it generalizes. In other words, once the optimal $f$ is found for the training set $(X,Y)$, and given another data set $(U,V)$, how close is $f(U)$ to $V$. 
	
	One of the most successful methods in tackling this kind of problems is Artificial Neural Networks (ANN), a subset of ML.  As the name suggests, an ANN is a set of connected building blocks called neurons which are meant to mimic the operations of biological neurons (\cite{anthony_bartlett_1999,Wang2003}).  Different kinds of ANNs can be built by varying the number of, connections between, and operations of individual neurons. The operations performed by these neurons depend on a number of parameters, called \textit{weights} and some nonlinear function, called the \textit{activation}. 
	At a high level, an ANN is just the function $f$ that was described earlier. Since the network architecture is chosen at the start, finding the optimal $f$ boils down to finding the optimal weight parameters that minimize the cost function $L$. 
	
	Regardless of the type of ANN used, the process of finding the optimal weights is more or less the same, and works as follows. After the network architecture is chosen, the weights are initialized, then a variant of gradient descent is applied to the training data. Gradient descent changes the parameters iteratively, at a certain rate proportional to their gradient, until the loss value is sufficiently small (\cite{GD}). The proportionality constant is called the \textit{learning rate}. While this process is well known, there is to date no clear prescription for choice of the different components . The main difficulty arises from the fact that the loss function contains multiple minima with different generalization properties. In other words, not all minima of the loss function are equal in terms of generalization. Which minimum is reached at the end of training phase depends on the initial values chosen for the weights, the optimization algorithm used, including the learning rate and the training dataset (\cite{Zhang2016}). In the absence of clear theoretical prescriptions for the components one has to rely on experience and best practices (\cite{Bengio2012}).
	
	One popular type of ANN is the feedforward network, where neurons are organized in layers, with the outputs of each layer \emph{fully} connected to the inputs of the next. By increasing the number of layers (whence the "deep" in "deep learning") many types of data can be modelled to a high degree of accuracy. Fully connected ANN, however, have some shortcomings, such as the large number of parameters, slow convergence, overfitting, and most importantly, failure to detect \emph{local} patterns. Almost all the aforementioned shortcomings are solved by using convolution layers.
	
	\begin{figure*}[!h]
		\centering
		\includegraphics[scale=0.48]{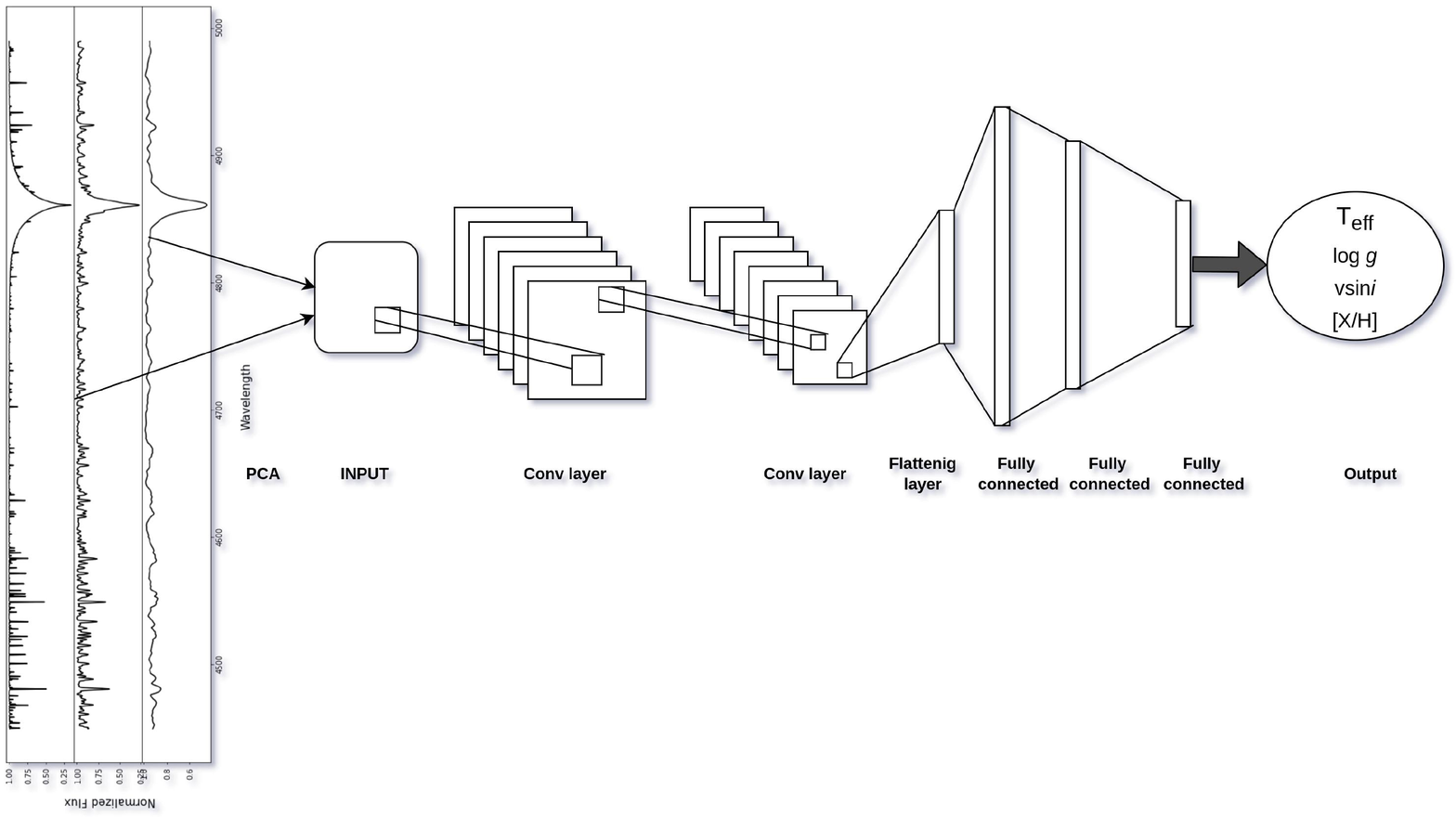}
		\caption{CNN architecture used in this work. A PCA dimension reduction transforms the spectra into a matrix of input coefficient. This input passes through several convolutional layers and fully connected layers in order to train the data and predict the stellar parameters.}
		\label{architecture_CNN}
	\end{figure*} 
	\subsection{Convolutional Neural Network}
	\label{CNN}
	A convolutional neural network (CNN) is a multi-layer network where at least one of the layers is a convolution layer \cite{lecun1989generalization}. As the name suggests, the output of a convolution layer is the result of a convolution operation, rather than matrix multiplication, as in feedforward layers, on its input. Typically, this convolution operation is performed via a set of filters. CNNs have been very successful in image recognition tasks (\cite{10.1007/978-3-319-16841-8_52}). Most commonly, CNNs are used in conjunction with pooling layers. In this work, since the input to the CNN has been already processed with PCA to reduce the dimension of the training database, we decided to omit pooling layers in our work. 
	Even though CNNs have been mostly used for processing image data, which can be viewed as  2-D grid data, they can also be used for 1-D data as well.
	

	The architecture of a CNN differs among various studies. There is no perfect model, it all depends on the type and size of the input data, and on the type of the predicted parameters. In this work, we will not be constraining the architecture of the model but rather we will be providing the best strategy to constrain the parameters of the model for a specific and defined architecture.
	Figure~\ref{architecture_CNN} shows a flow-chart of a typical CNN.  Table~\ref{tab:cnn} represents the different layers, the output shape for each layer, and the number of parameters used in our model. In the same table, "Conv" stands for convolutional layer, "Flat" for flattening layer which transform the matrix of data to one dimensional, and "Full" stands for fully connected layer. The total numbers of parameters to be trained every iteration is 764\,357. The choice of such an architecture is based on aF trial and error procedure that we performed in order to find the best model that can handle all types of training databases used in this work. The strategy of selecting the number of hidden layers and the size of the convolution layers will be described in a future paper. We decided to do all our tests using the ML platform \texttt{TensorFlow}\footnote{\url{https://www.tensorflow.org/}} with the \texttt{Keras}\footnote{\url{https://keras.io/}} interface. The reason is that these two options are open-source and written in \texttt{Python}.
	
	Although the calculation time is an important parameter constraining the choice of a network, we have decided not to take it into consideration while selecting the optimal Network. The reason for that is that the calculation time depends mainly on the network's architecture which is not discussed in this paper. Two parameters are also constraining the calculation time, the number of epochs and the batch size (related to the size of the TDB). Calculation time increases with increasing epochs number and decreases with increasing batches size. The main goal of this work is to find the optimal configuration for the parameters independently of the calculation time and the Network's architecture. As a rule of thumb, using a Database of 70\,000 spectra and 50 eigenvectors, it takes around 17 hours to run the CNN over 2000 epochs using 64 batches and a Dropout of 30\%. These calculations are done on a Intel Core i7-8750H CPU \@ 2.20GHz $\times$ 6 CPU.

	\begin{table}
		\centering
		\begin{tabular}{||c|cc||}
			\hline
			Layer & Output shape & \# Parameters \\ 
			\hline
			Conv & 50 $\times$ 8&             40        \\
			Conv & 50 $\times$ 4&             132        \\
			Conv & 50 $\times$ 4&             68        \\
			Flat & 200 & 0 \\
			Full & 1\,024 & 205\,824 \\
			Full & 512 & 524\,800 \\
			Full & 64 & 32\,832 \\
			Full & 10 & 650 \\
			Full & 1 & 11 \\
			\hline 
		\end{tabular}
		\caption{Different layers that are used in the CNN used in our work. }
		\label{tab:cnn}
	\end{table}


	\subsubsection{Data Augmentation}
	\label{data_aug}
	Data augmentation is a regularization technique that increases the diversity of the training data by applying different transformations to the existing one. It is usually used for images classification (\cite{Shorten2019ASO}) and speech recognition (\cite{Jaitly2013VocalTL}). We tested this approach in our procedure in order to take into account some modifications that could occur in the shape of the observed spectra due to a bad normalisation or inappropriate data reduction. We also took into account the fact that observed spectra are affected by noise and that the learning process should include the effect of this parameter.\\
	
	For each spectrum in the TDB, 5 replicas were performed. Each of these 5 replicas has different amount of flux values but they all have the same stellar labels \Teff, \logg, \met, and \vsini. The modifications are done as follows:
	\begin{itemize}
		\item A Gaussian noise is added to the spectrum with a SNR ranging randomly between 5 and 300.
		\item The flux is multiplied in a uniform way with a scaling factor between 0.95 and 1.05.
		\item The flux is multiplied with a new scaling factor and noise was added.
		\item The flux is multiplied by a second degree polynomial with values ranging between 0.95 and 1.05 and a having its maximum randomly selected between 4\,450 and 5\,000 \AA.
		\item The flux is multiplied by a second degree polynomial and Gaussian noise added to it.
	\end{itemize}
	
\begin{figure*}[!h]
		\centering
		\includegraphics[scale=0.55]{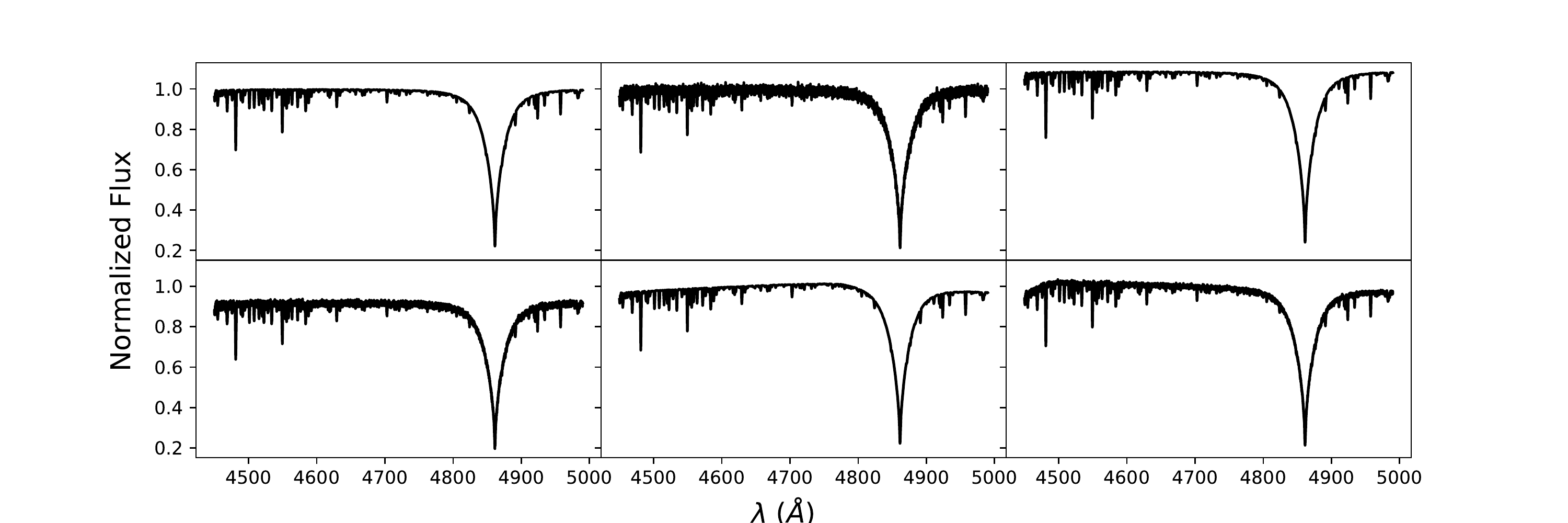}
		\caption{The effect of the data augmentation on the shape of the spectra.  \textit{Upper left}: spectrum represents displays the original synthetic spectra. \textit{Upper middle}: Gaussian noise added to the synthetic spectra. \textit{Upper right}: synthetic spectrum with the intensities multiplied by a constant scale factor. \textit{Bottom left}: Gaussian noise added to the synthetic spectra and multiplied by a constant scale factor. \textit{Bottom middle}: synthetic spectrum with the intensities multiplied by a second degree polynomial. \textit{Bottom right}: Gaussian noise added to the synthetic spectra and multiplied by a second degree polynomial. All these spectra have the same stellar parameters (\Teff=8800 K, \logg=4.3 dex, \vsini=45 \kms, \met=0.0 dex). }
		\label{data_aug_fig}
	\end{figure*}

	The purpose of this choice is to increase the dimension of the TDB from $N_{\mathrm{spectra}} \times N_{\lambda}$ to $6\times N_{\mathrm{spectra}} \times N_{\lambda}$ and to introduce some modifications in the training spectra that could appear in the observations that we need to analyze. Such modifications are the noise and the commonly observed departures from a perfect continuum normalization. Distorsions in observed spectra could appear due to bad selection in the continuum points.
	We have tested the two options, with and without data augmentation and the results are shown in Sec.~\ref{results}. Figure.~\ref{data_aug_fig} displays one synthetic spectrum having \Teff=8800 K, \logg=4.0 dex, \vsini=14 \kms, and \met=0.0 dex as well as the extra 5 modifications that were performed on this spectrum. We have decided to use a continuous SNR between 5 and 300 but different modifications could be tested. As an example, \cite{2017MNRAS.465.4556G} adapted the SNR of the spectra used in the training dataset to the SNR of the spectra for which the atmospheric parameters are needed (evaluation set). They concluded, that in case of \Teff, only two regression models are needed (SNR = 50 and 10) to cover the entire SNR range.

	\subsubsection{Initialisers: Kernel and Bias}
	The initialization defines the way to set the initial weights. There are various ways to initialize, and we will be testing the following:
	\begin{itemize}
		\item \textit{Zeros}: weights are initialized with 0. In that case, the activation in all neurons is the same and the derivative of the loss function is similar for every weight in every neuron.  This results in a linear behavior for the model.
		\item \textit{Ones}: a similar behavior as the Zeros but using the value of 1 instead of 0.
		\item \textit{RandomNormal}: initialization with a normal distribution.
		\item \textit{RandomUniform}: initialization with a uniform distribution.
		\item \textit{TruncatedNormal}: initialization with a truncated normal distribution.
		\item \textit{VarianceScaling}: initialization that adapts its scale to the shape of weights.
		\item \textit{Orthogonal}: initialization that generates a random orthogonal matrix.
		\item \textit{Identity}: initialization that generates the identity matrix.
		\item \textit{Lecun\_uniform}: LeCun uniform initializer (\cite{726791}).
		\item \textit{Glorot\_normal}: Xavier normal initializer (\cite{pmlr-v9-glorot10a}).
		\item \textit{Glorot\_uniform}: Xavier uniform initializer (\cite{pmlr-v9-glorot10a}).
		\item \textit{he\_normal}: He normal initializer (\cite{He:2015dtg}).
		\item \textit{Lecun\_normal}: LeCun normal initializer (\cite{726791}).
		\item \textit{he\_uniform}: He uniform variance scaling initializer (\cite{He:2015dtg}).
	\end{itemize}
	
	For all of these initializers, the biases are initialized with a value of zero. It will be shown later that most of these initialisers give the same accuracy except for the Zeros and Ones. 
	
	
	\subsubsection{Optimizer} 
	Once the (parameterized) network architecture is chosen, the next step is to find the optimal values for the parameters. If we denote by $\theta$ the collective set of parameters, then, by definition, the optimal values, $\theta_*$, are the ones that minimize a certain loss function $L(\theta)$; a measure of difference between the predicted and the actual values. This optimization problem is, typically, solved in an iterative manner, by computing the gradient of the loss function with respect to the parameters.
	
	Let $\theta_t$ denote the set of parameters at iteration \textit{t}. The iterative optimization process produces a sequence of values, $\theta_1,\ldots,\theta_*$ that converges to the optimal values $\theta_*$. At a given step $t$ we define the history of that process as the set $\mathcal{H}_t=\{\theta_i,L(\theta_i),\nabla L(\theta_i)\}_{i=0}^t$. The values $\theta_{t+1}$ are obtained from $\theta_t$ according to some update rule $\mathcal{U}$
	\begin{align}
		\theta_{t+1}=\mathcal{U}(H_t,\gamma_t)
	\end{align}
	where $\gamma_t$ is a set of hyperparameters such as the learning rate. 
	
	Different optimization techniques use a different update rule. For example, in the so-called "vanilla" gradient descent, the update rule depends on the most recent gradient only:
	\begin{align}
		\theta_{t+1}=\theta_t-\gamma\nabla L(\theta_t)
	\end{align}
	Other methods,  include the whole history with different functional dependence on the gradient and different rates for each step (see \cite{choi2020empirical} for a survey).
	Different optimization techniques are available in keras and we will be testing the following:
	\begin{itemize}
		\item \textit{Adam}: an Adaptive moment estimation that is widely used for problems with noise and sparse gradients. Practically, this optimizer requires little tuning for different problems.
		\item \textit{RMSprop}: a Root Mean Square propagation that iteratively updates the learning rates for each trainable parameter by using the running average of the squares of previous gradients. 
		\item \textit{Adadelta}: it is an adaptive delta, where delta refers to the difference between the current weight and the newly updated weight. It also works as a stochastic gradient descent-method.
		\item \textit{Adamax}: an adaptive stochastic gradient descent method, and a variant of Adam and is based on the infinity norm. It is also less sensitive to the learning rates than other optimizers.
		\item \textit{Nadam}: Nesterov-accelerated Adam optimizer that is used for gradients with noise or with high curvatures. It uses an accelerated learning process by summing up the exponential decay of the moving averages for the previous and current gradient. It is also an adaptive learning rate algorithm and requires less tuning of the hyperparameters.
	\end{itemize}

	\subsubsection{Learning Rate}
	
	As mentioned in the beginning of the section the training rate can affect the minimum reached by the loss function and therefore has a large effect on the generalization property of the solution. In this paper, we followed the recommendation of (\cite{Bengio2012}) and chose the learning rate value to be half of the largest rate that causes divergence.
	
	\subsubsection{Dropout}
	\label{drop}
	Dropout is a regularization technique for neural networks and deep learning models that prevents the network from overfitting (\cite{srivastava2014dropout}). When Dropout is applied, randomly selected neurons are removed each iteration of the training and do not contribute to the forward propagation and no weight updates are applied to these neurons during backward propagation. Statistically this has the effect of doing ensemble average over different sub-networks obtained from the original base network. 
	We tried to find the optimal number for the dropped out fraction of neurons. Dropout layers are put after each convolutional ones. Tests were performed with Dropout fraction ranging between 0 and 1.

	\subsubsection{Pooling}
	Pooling layers is a way to down sample the features (i.e. reducing the dimension of the data) in the database by taking patches together during the training. The most common pooling methods are the average and the max pooling  \cite{zhou1988computation}. The average one summarizes the mean intensity of the features in a patch and the max one considers only the most intense (\textit{ie}. highest value) value in a patch. The size of the patches and the number of filters used is decided by the user. The standard way to do that is to add a pooling layer after the convolutional layer and this can be repeated one or more times in a given CNN. However, pooling makes the input invariant to small translations. In image detection, we need to know if the features exist and not their exact position. That is why this techniques has shown to be valuable when analyzing images (\cite{Goodfellow-et-al-2016}). This is not the case in spectra because the position of the lines needs to be well known (see Sec.~\ref{results}). But also, as discussed previously, pooling layers are not needed in our case because the dimension of the TDB was already reduced drastically by applying PCA.

	\subsubsection{Activation Functions}
	The activation function is a non linear transformation that is applied on the output of a layer and this output is then sent to the next layer of neurons as input. Activation functions play a crucial role in deriving the output of a model, determining its accuracy and computational efficiency. In some cases, activation functions might prevent the network from converging.\\

	The activation function for the \textit{inner} layers of a deep networks must be nonlinear, otherwise no matter how deep the network is, it would be equivalent to a single layer (i.e. regression/logistic regression). Having said that, we have tested 5 activation functions that are:
	\begin{itemize}
		\item sigmoid: $f(x)=\dfrac{1}{1+e^{-x}}$\\
		\item tanh: $f(x)=\dfrac{e^{x}-e^{-x}}{e^{x}+e^{-x}}$\\
		\item relu: $f(x)=Max(0,x)$\\
		\item elu: $f(x) = \left\{
		\begin{array}{ll}
			x & x\geq 0 \\
			\alpha (e^x -1) & x< 0
		\end{array}
		\right.$\\
		\item selu: $f(x)= \left\{
		\begin{array}{ll}
			\lambda x & x\geq 0 \\
			\lambda \alpha (e^x -1) & x< 0
		\end{array}
		\right.$

	\end{itemize}
	
	It is important to note that in this section we discuss the choice of the activation function for inner layers only. The choice of the activation for the last layer is usually more or less fixed by the type of the problem and how one is modelling it. For example, if one is performing binary classification then a sigmoid like activation is usually used (or softmax for multiclass classification) and interpreted as a probability. Whereas, for regression like problems a linear activation is usually used for the last layer. In our case, which is a purely regression problem, the last layer will have a linear activation function.
	
	The sigmoid and tanh restrict the magnitude of the output of the layer to be $\le 1$. Both, however, suffer from the \textit{vanishing gradient} problem (\cite{glorot2011}). For relatively large magnitudes both functions \textit{saturate} and their gradient becomes very small. Since deep networks rely on backpropagation for training the gradient, the first few layers, being a product of the succeeding layers, become increasingly small. The rectifier class of activation, relu, elu, etc... seem to minimize the vanishing gradient problem. Also, they lead to sparse representation which seems to give better results (\cite{He2015,maasrectifier}).

	\subsubsection{Loss Functions}
	The Loss Function controls the prediction error of a NN as explained in Sec.~\ref{DL}. It is an important criterium in controlling the updates of the weights in a NN, mainly during the backward propagation. The selection of the type of the loss function is decided depending on the types of output labels. If the output is a categorical variable, one can use the Categorical Crossentropy or the Sparse Categorical Crossentropy. If we are dealing with a binary classification, Binary Crossentropy will be the normal choice for a Loss Function. Finally, in case of a regression problem like the one used in stellar spectra parameters determination, variants of Mean Squared Error Loss functions are used. In our work, we have tested the following functions:\\
	
	\begin{itemize}
		\item \textit{Mean Squared Error}: $\frac{1}{N}\sum_{i=1}^{N}(y_i - \hat{y}_i)^2$ \\
		
		\item \textit{Mean Squared Logarithmic Error}:$\frac{1}{N}\sum_{i=1}^{N}(\log \frac{1+y_i}{1+\hat{y}_i})^2$ \\
		\item \textit{Mean Absolute Error}:  $\frac{1}{N} \sum_{i=1}^{N} |y_i - \hat{y}_i|$
	\end{itemize}
	$y$ being the actual label, $\hat{y}$ the predicted ones, and $N$ the number of spectra in the training dataset.\\
	Loss functions selection can differ from one study to the other (\cite{loss}). For that reason we have tested the above three functions in deriving the stellar parameters.

	\subsubsection{Epochs}
	The number of Epochs is the number of times the whole dataset is used for the forward and the backward propagation. The number of Epochs controls the number of times the weights of the neurons are updated. While increasing the number of Epochs, we can move from underfitting to overfitting passing through the optimal solution for our network. 
	
	\subsubsection{Batches}
	Instead of passing the whole training dataset into the NN, we can divide it in $N_{\rm{Batches}}$ batches and iterate on all batches per epoch. In that case, the number of iterations will be the number of batches needed to complete one epoch. Batches are used in order to avoid the saturation of the computer memory and the decrease of iterations speed. However, the selection of the optimal batch number is not straightforward. Adopted values are usually 32, 64 or 128 (\cite{Keskar2016}). 
	
	One of the most important measures of the success for a deep neural network is how well it generalizes on some test data, not included in the training phase. In current deep neural networks the loss function has multiple minima. Many experimental studies have shown that, during the training phase, \textit{the path to reaching} a minimum  is as important as the final value (\cite{Neyshabur2017,Zou2020,Zhang2016}). A good rule of thumb is that a "small", less than 1\% the size of the data, batch size generalize better than "large" batches, about 10\% of the training data (\cite{Keskar2016}).

	\section{Results and Analysis}
	\label{results}
	The effect of each CNN parameter on the accuracy of the stellar parameters has been tested. To do so, we have used the same CNN with the same parameters for all our tests while changing only the concerned one at each time. For example, to find the best epoch numbers, we fix the Activation function, the optimizer, the number of Batches, the Dropout percentage, the loss function and the kernel initialiser while iterating on the number of epochs. The same parameters are used again for finding the optimal dropout percentage and so on. The fixed values used in these calculations are the he\_normal for the kernel initialiser, the mean squared error for the loss function, the "ADAM" optimizer, the relu activation function, 50\% of Dropout, 64 Batches. These tests are performed with epochs of 100, 500, 1000, 2000, 3000, 4000 and 5000. In all tests, the distribution of Training and Validation are 80\% and 20\% respectively.
	
	The results will be a combination of Test errors spanning over different number of epochs for each stellar parameter and CNN configuration. The variation with the number of epochs ensures that the trends are real and not due to local minima as a result of the low number of iterations. The Tests are a collection of 110\,000 synthetic spectra, half of them without noise and half with random noise as introduced in Sec.~\ref{LDB}.

	To better visualise the results and to have a better conclusion about the optimal configurations, we display in Figs.~\ref{fig:Teff_tests} to \ref{fig:Vrot_test} the relative error of the observations. These errors are calculated by dividing the values by the maximum observation standard deviation in all configurations (ie. including all epochs simulations). This will allow us to target the minimum values and pinpoint the best parameters.
	
	In what follows, we show the results that were performed using a training dataset of ~40\,000 randomly generated synthetic spectra in the ranges of Tab.~\ref{table_1}. In Sec.~\ref{small-big}, we discuss the effect of using a small or a large training database and the effect of using or not Data Augmentation.

	\subsection{Effective Temperature}
	\label{teff}
\begin{figure*}[!h]
		\centering
		\includegraphics[scale=0.35]{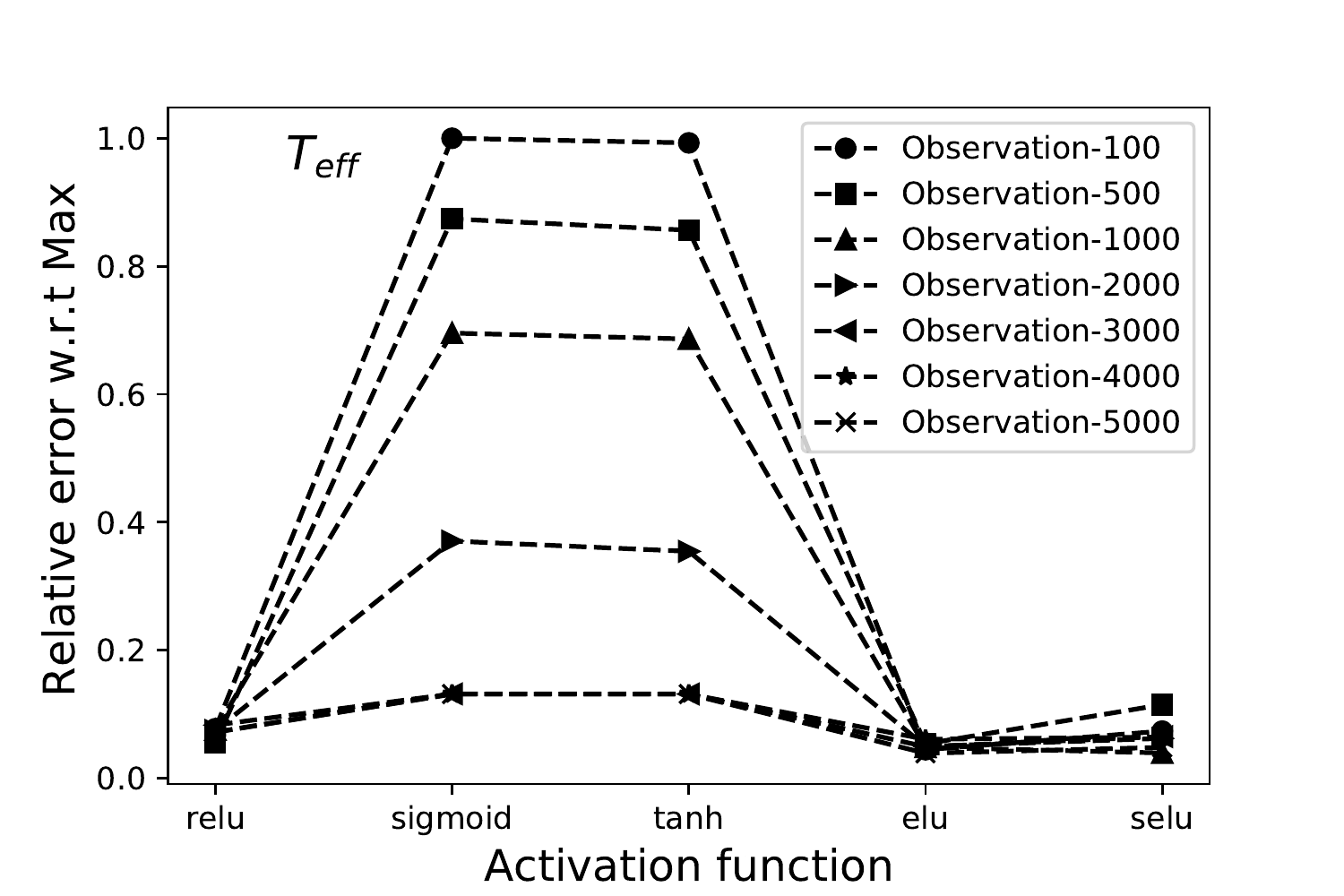}
		\includegraphics[scale=0.35]{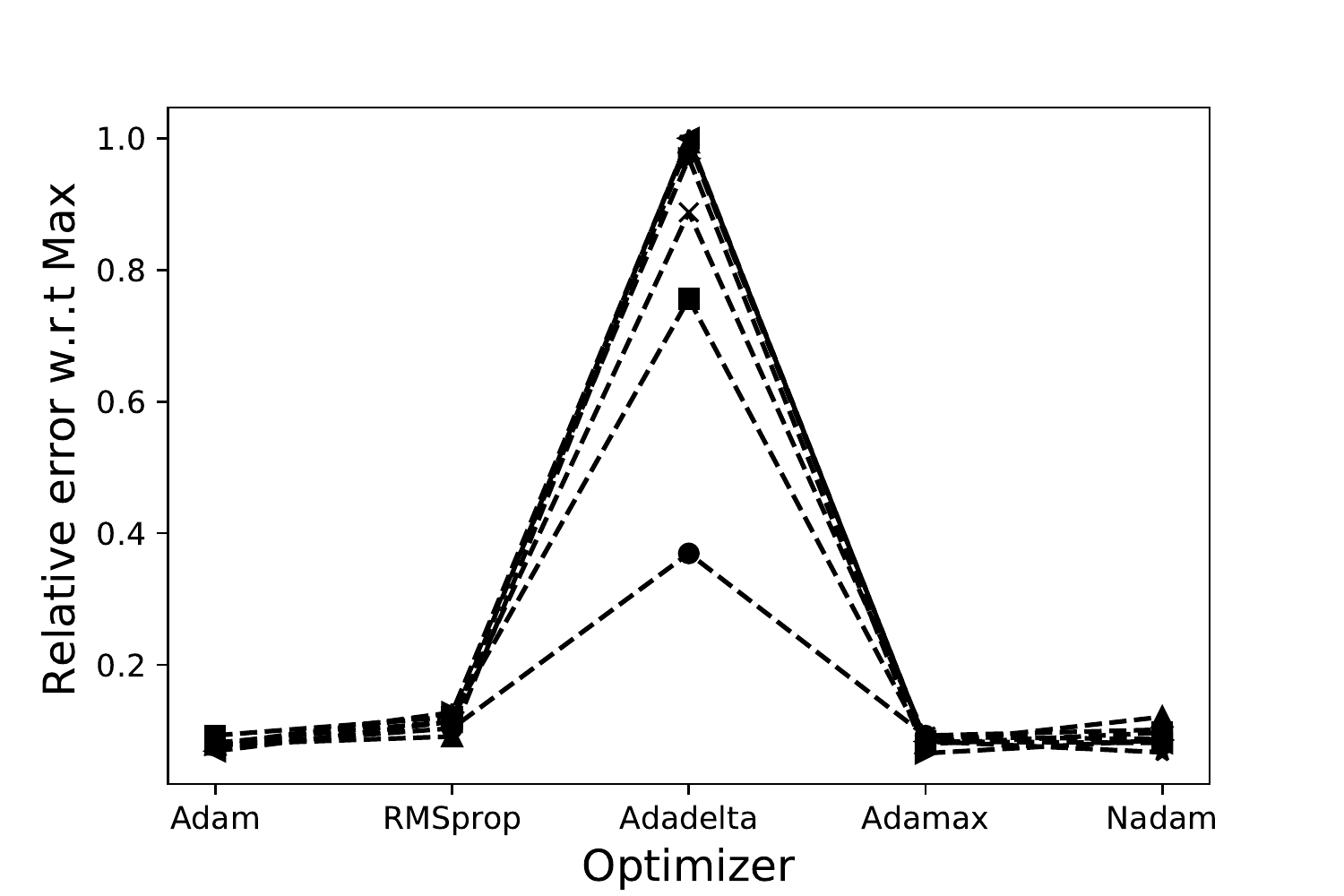}
		\includegraphics[scale=0.35]{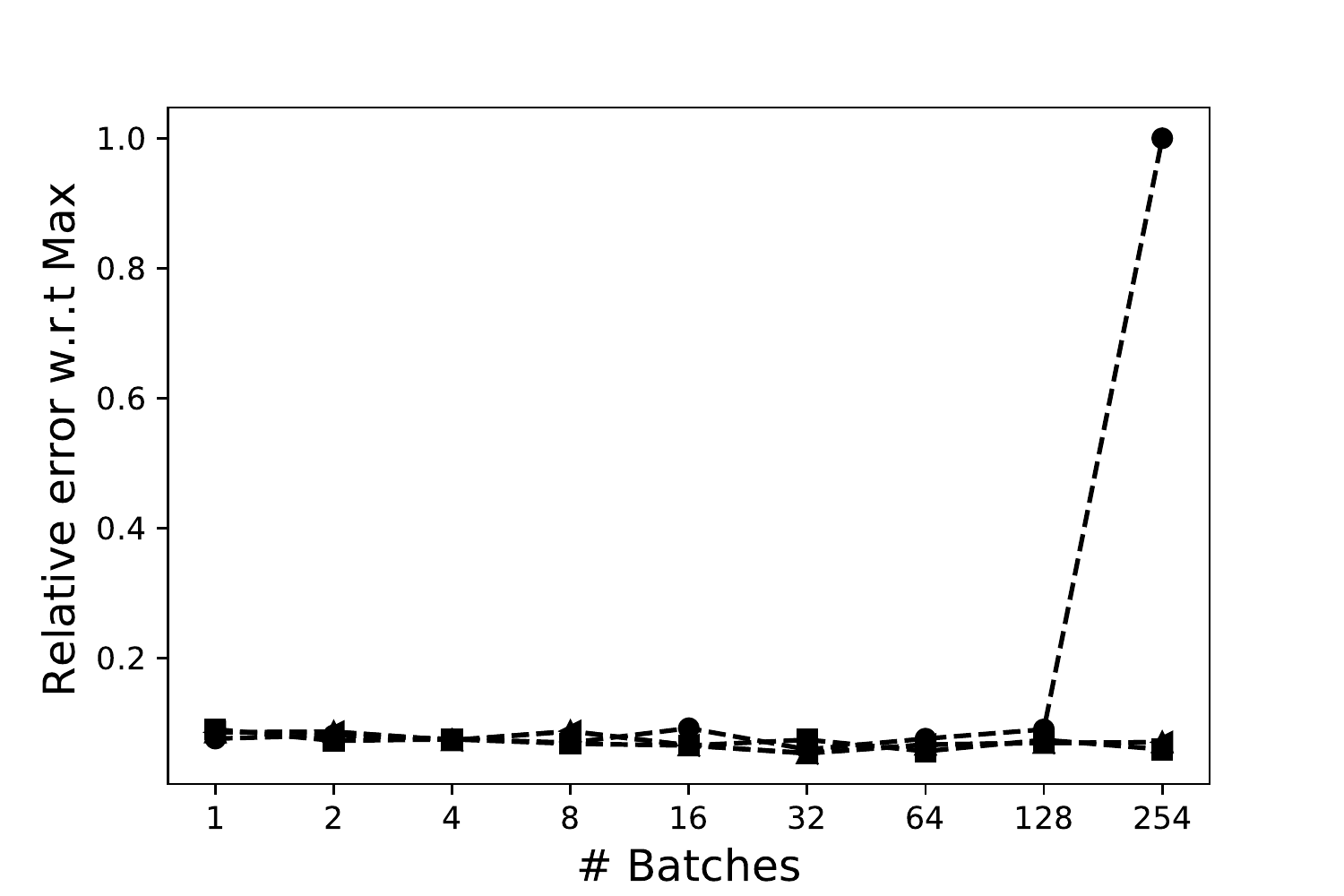}\\
		\includegraphics[scale=0.35]{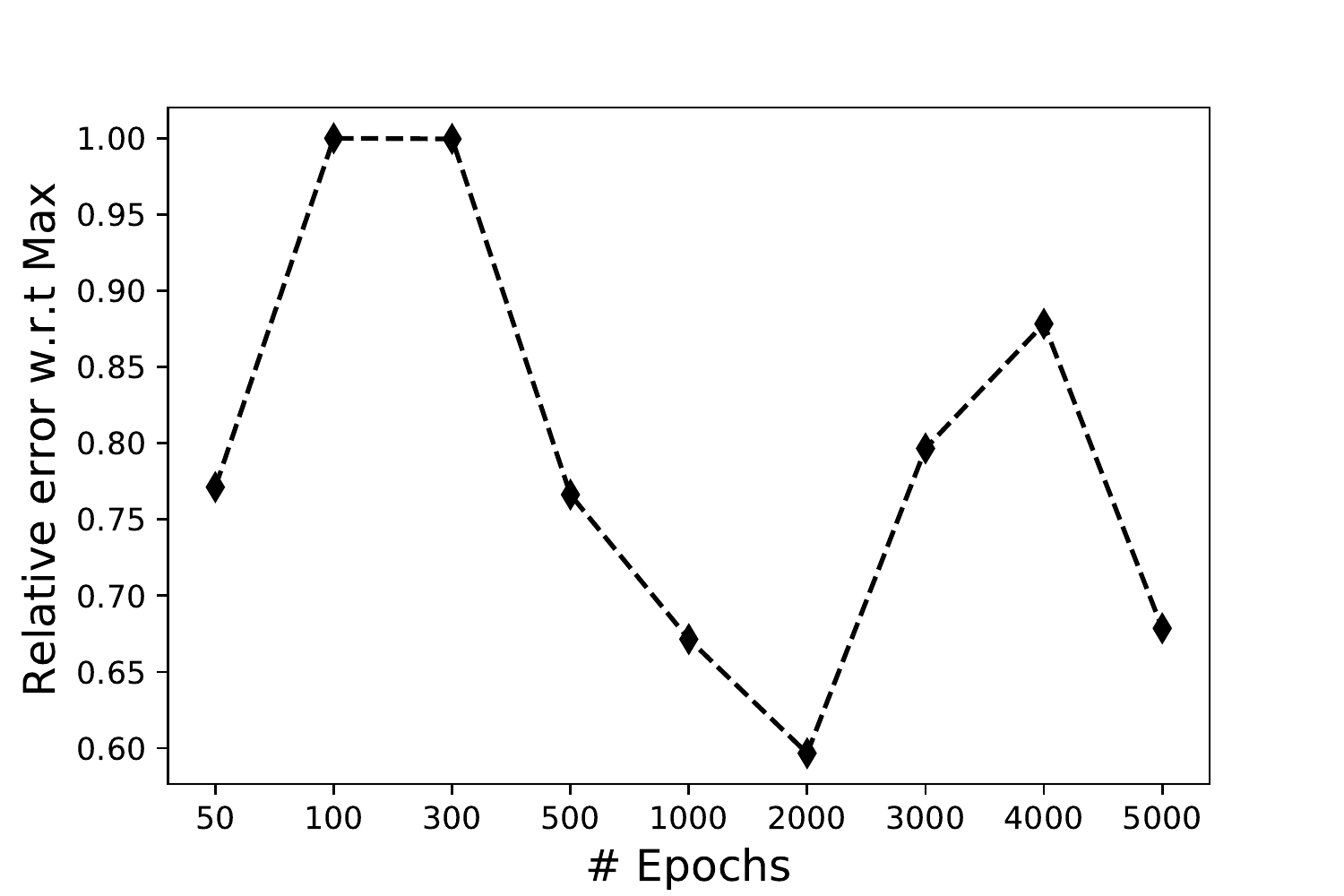}
		\includegraphics[scale=0.35]{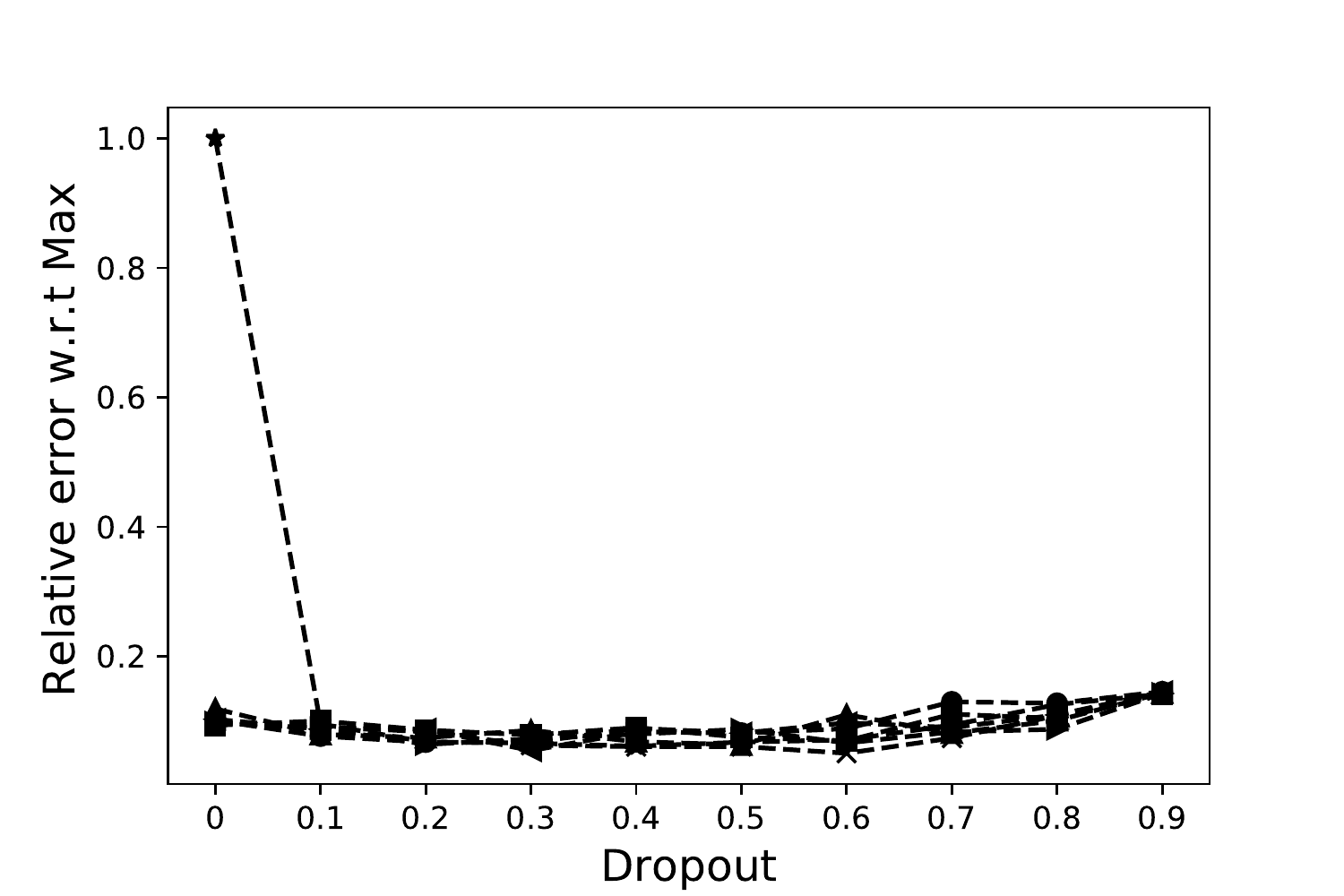}
		\includegraphics[scale=0.35]{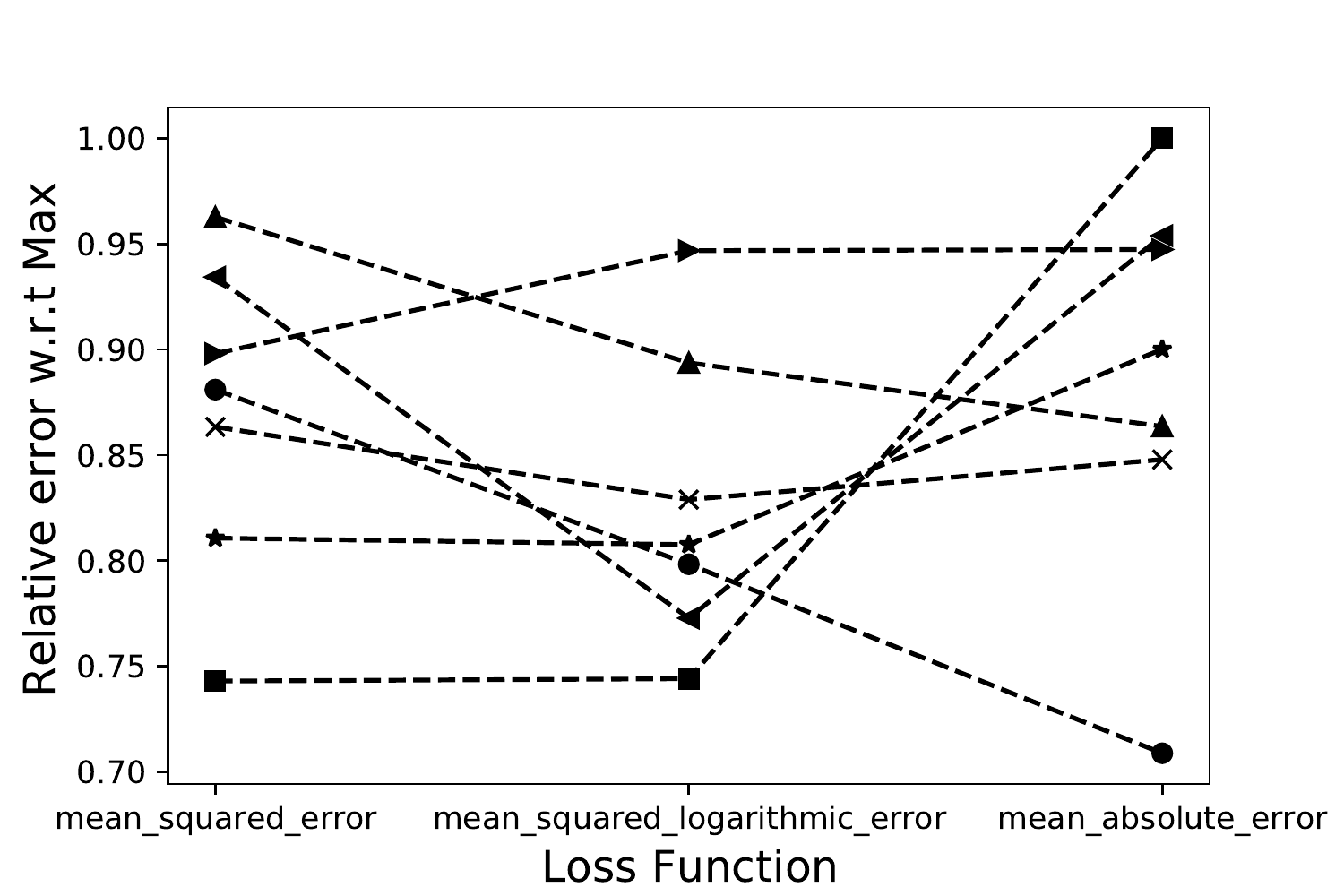}\\
		\includegraphics[scale=0.35]{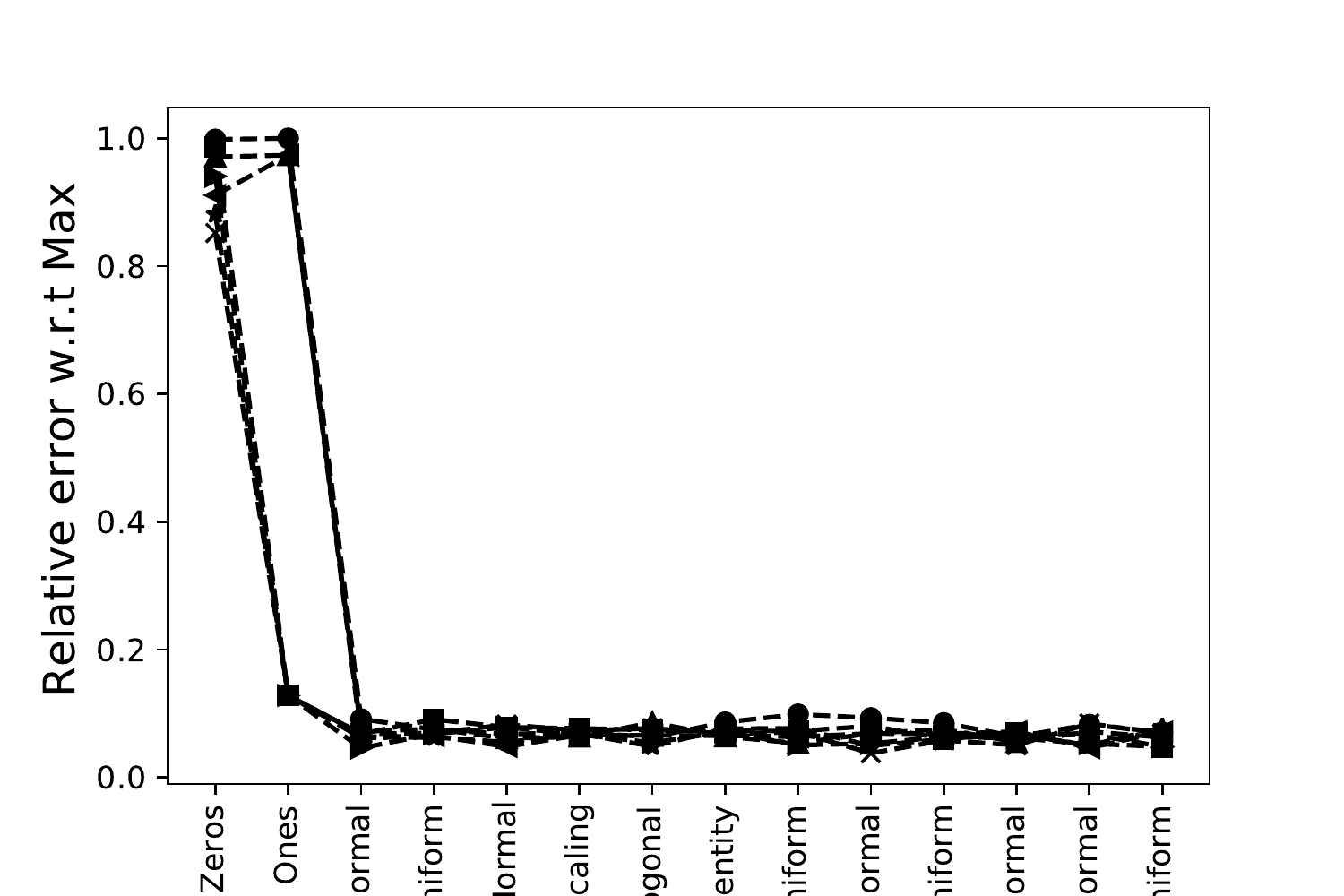}
		
		\caption{Effect of varying the CNN parameters on the accuracy of \Teff\ for different epoch numbers. The results are displayed by dividing the observation standard deviation by their maximum value in each test.}
		\label{fig:Teff_tests}
	\end{figure*}

	According to Fig.~\ref{fig:Teff_tests}, the use of a relu or elu activation functions leads to a similar conclusion within a difference of few percents. And this could be applied independently of the number of epochs. As for the Optimizer, Adam and Adamax optimizers seem to be consistently accurate across all epochs number. The optimal number of Batches is found to be between 32 and 64. The number of epochs is tightly related to the Batches number, however, in case of 64 Batches, the optimal number of epochs is found to be 2000. The dropout factor is, as introduced in Sec.~\ref{drop}, a regularization technique that avoids overfitting. This means that the optimal value depends on the size of the training database. In the case of our 40\,000 sample database, the optimal dropout is found to be between 10 and 60\%. Neural Networks minimize a loss function and accordingly derive the coefficient that will be used later to predict the parameters of the observations. Among the 3 loss functions that we tested, small differences are found among them. We will be using the mean squared logarithmic error for \Teff.  Finally, the initialisation of the network coefficients could be done using any initialiser with a exception of Zeros and Ones. Neural networks tend to get stuck in local minima when using these two options. The ratio of the standard deviation with respect to the maximum exhibits an up and down variation that resembles a jig-saw pattern with respect to the epochs number. This is mainly due to the fact that points correspond to different runs. Also we can notice that the variation of the relative error correspond to the smallest variations between the different hyperparameters, and this is the case for all stellar parameters. Of course the search will depend on the size of the training database, the spectral region, the spectral type, the resolution $\ldots$
	
	The optimal configuration that we found for \Teff\ corresponds to the following parameters:
	\begin{itemize}
		\item[] Activation function: relu.
		\item[] Optimizer: Adam.
		\item[] Batches: 64.
		\item[] Epochs: 2000.
		\item[] Dropout: 30\%.
		\item[] Loss function: mean squared logarithmic error. 
		\item[] Kernel initialiser: he\_normal. 
	\end{itemize}
	

	\subsection{Surface Gravity} 
	\label{logg}
\begin{figure*}[!h]
		\centering
		\includegraphics[scale=0.35]{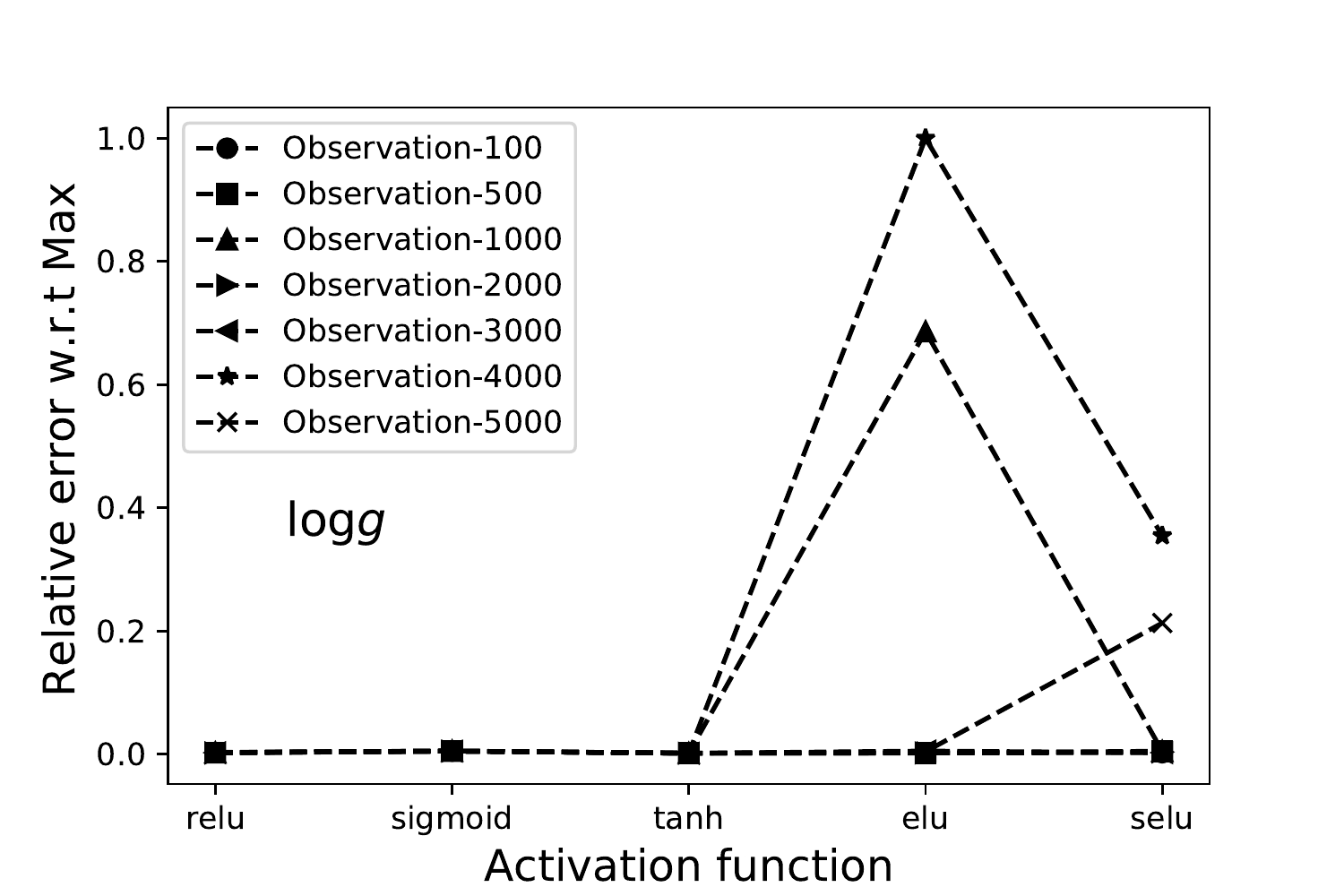}
		\includegraphics[scale=0.35]{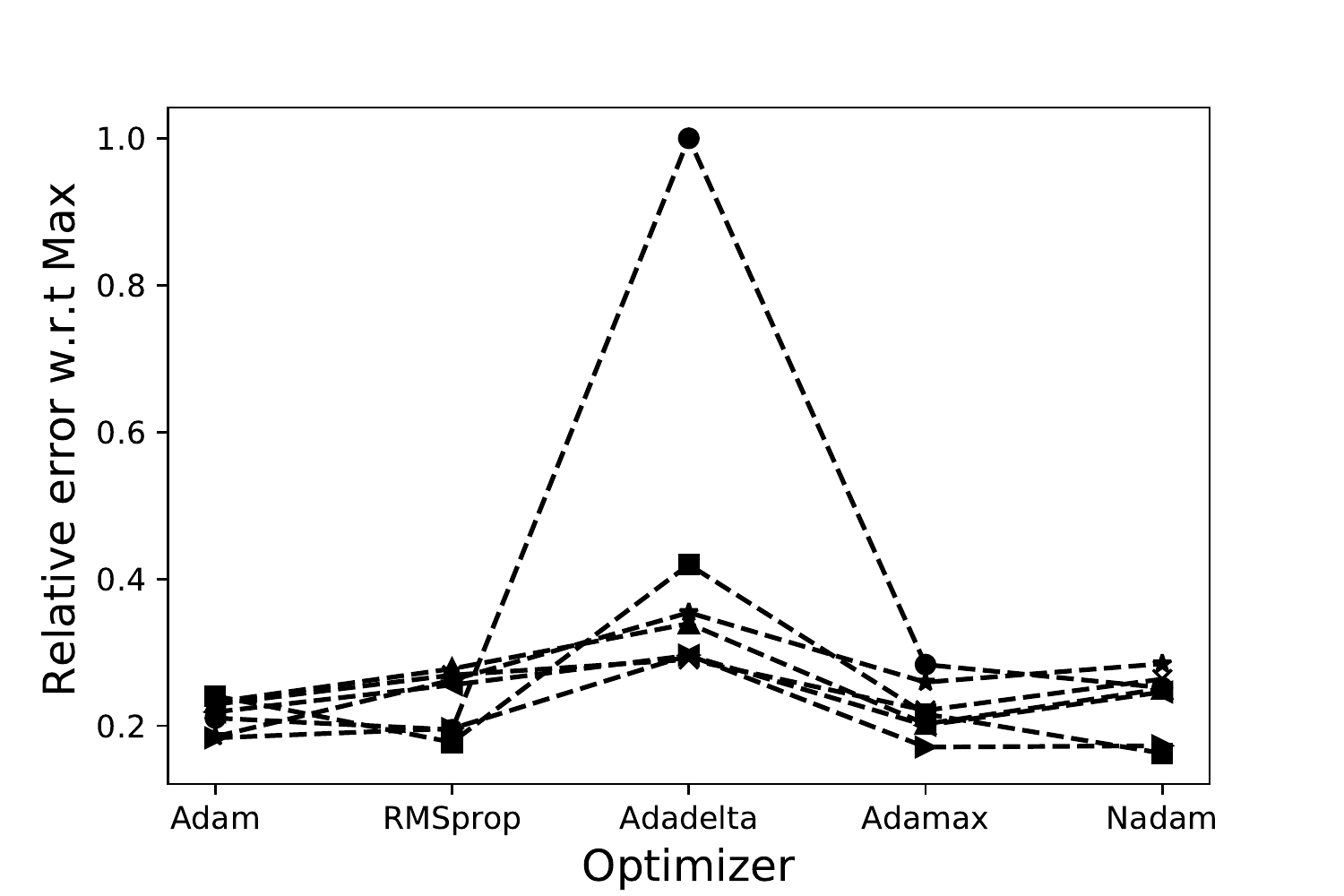}
		\includegraphics[scale=0.35]{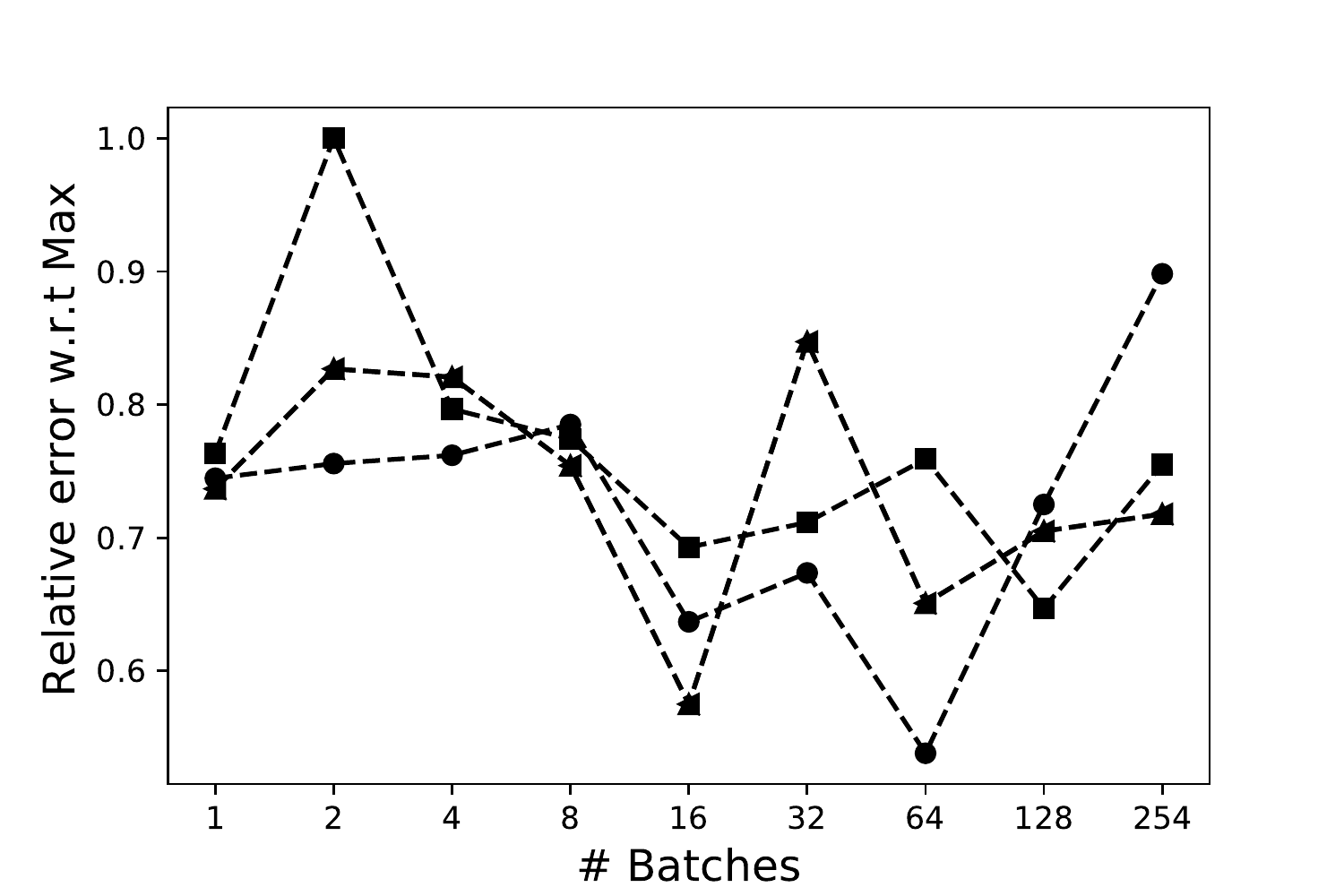}\\
		\includegraphics[scale=0.35]{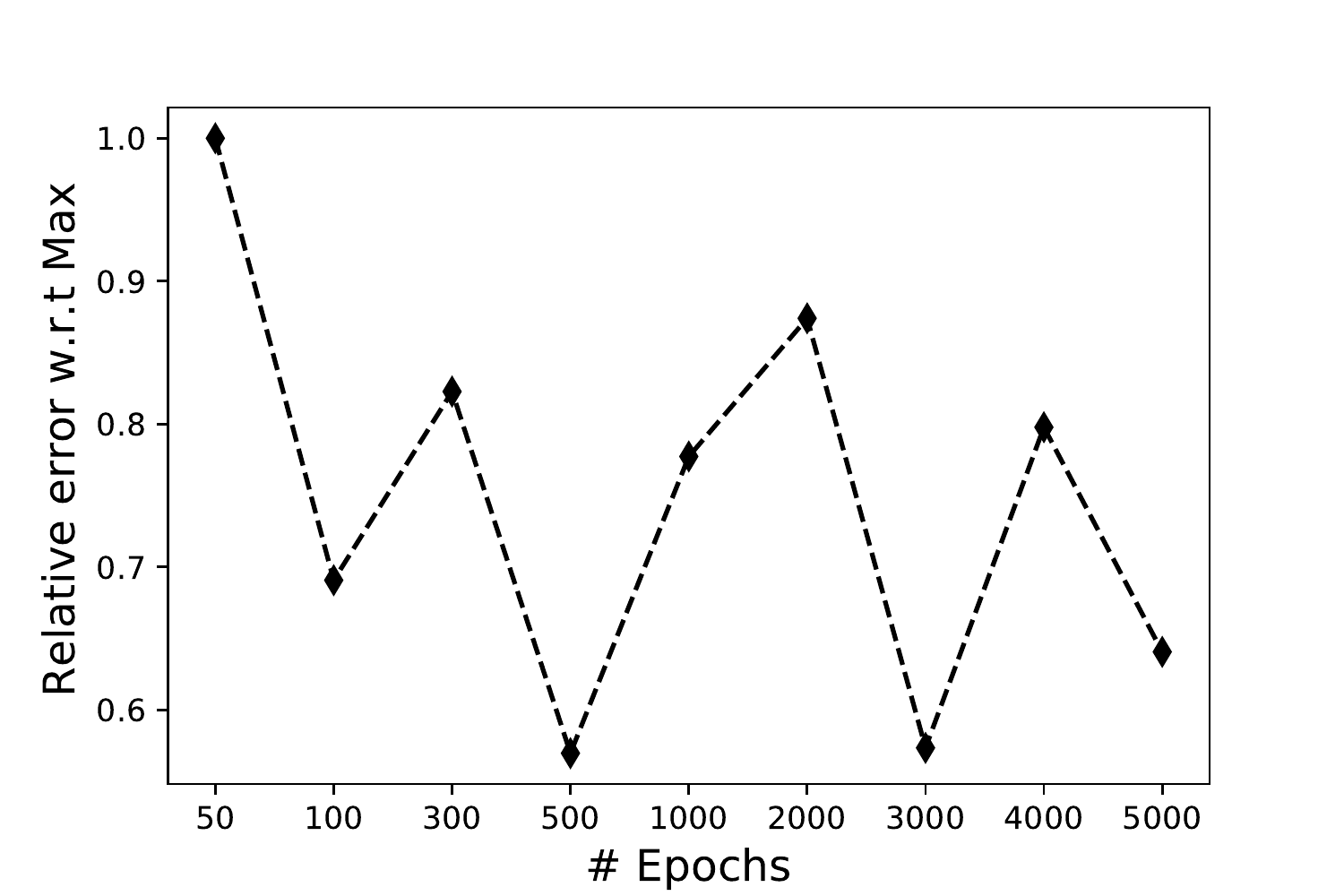}
		\includegraphics[scale=0.35]{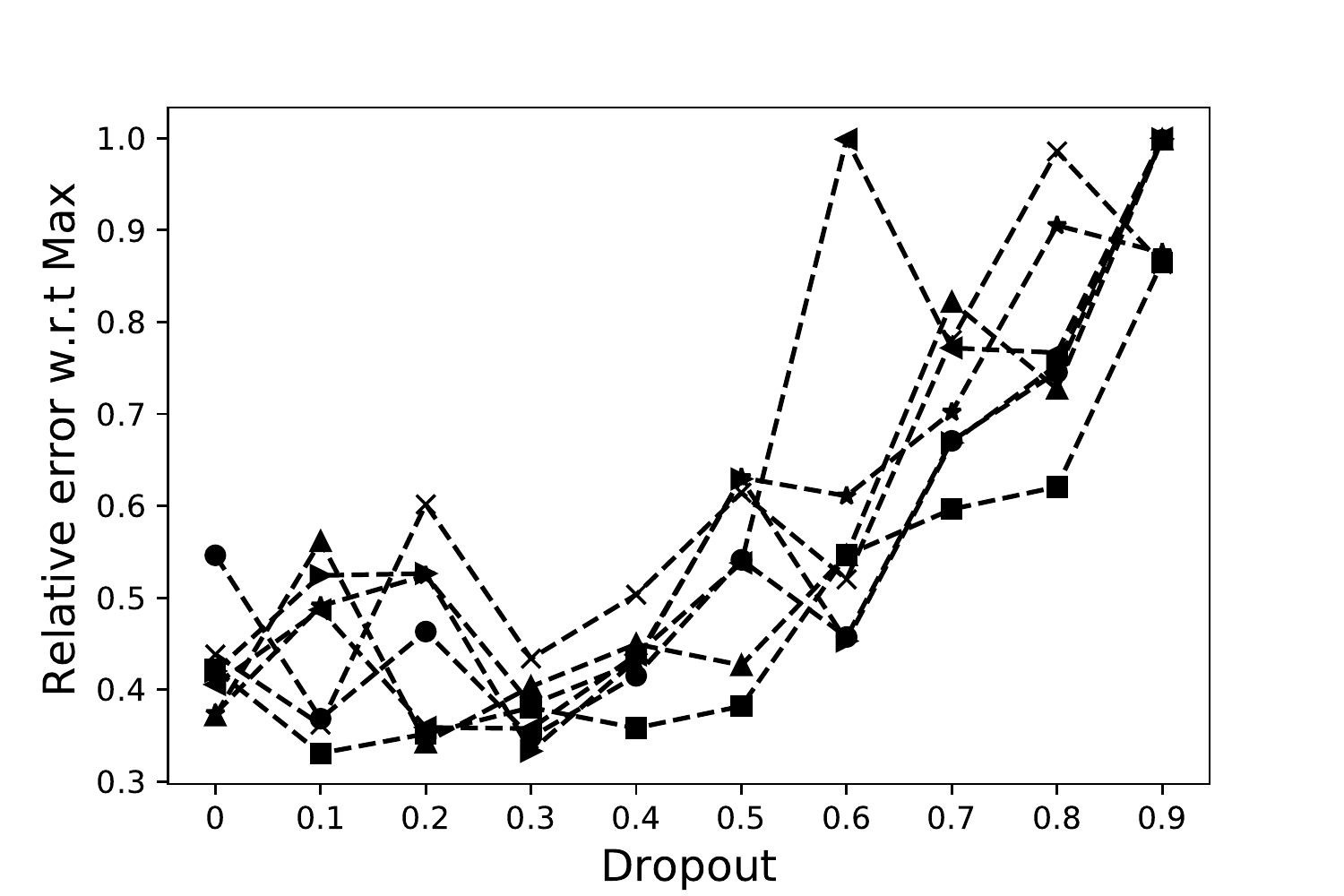}
		\includegraphics[scale=0.35]{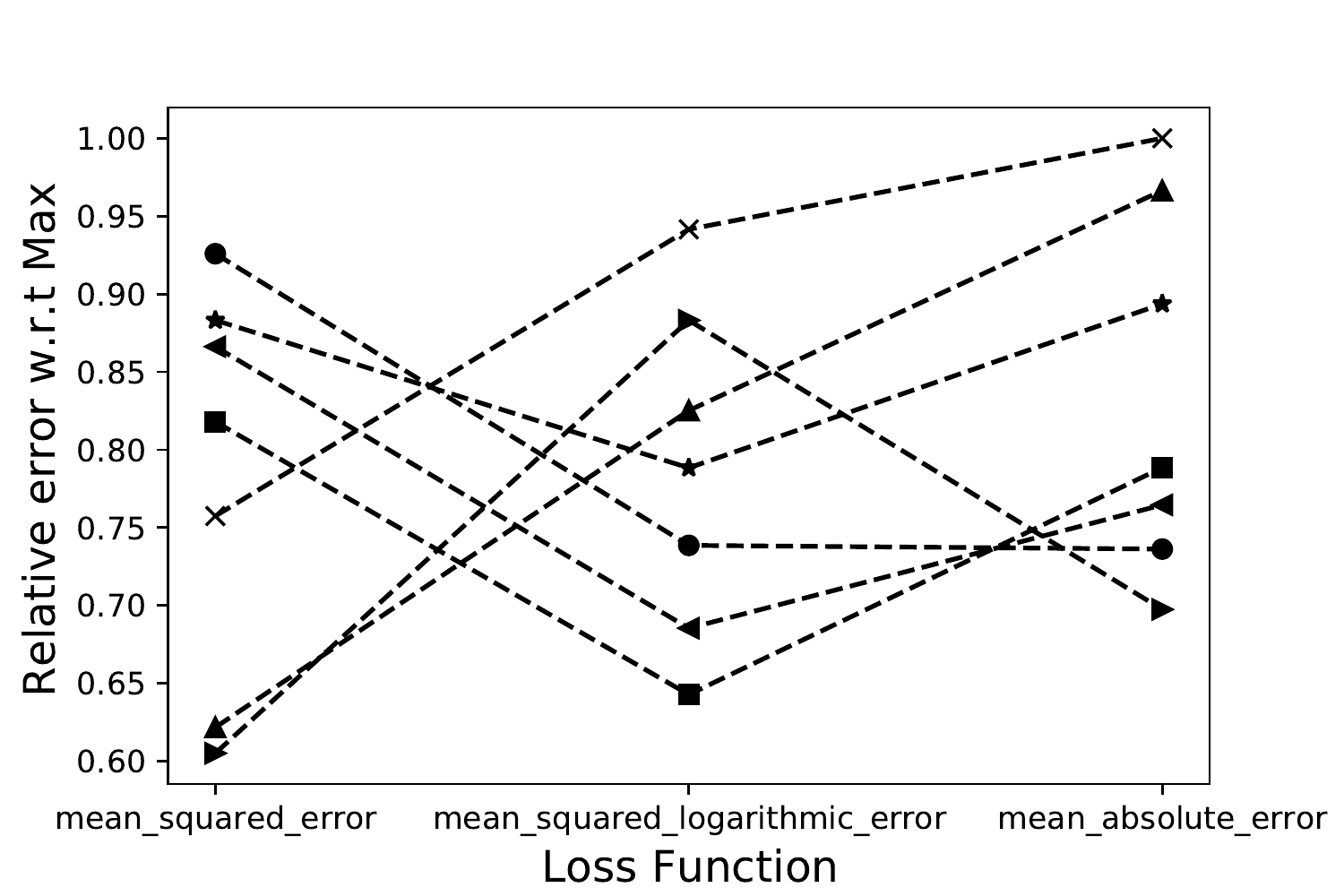}\\
		\includegraphics[scale=0.35]{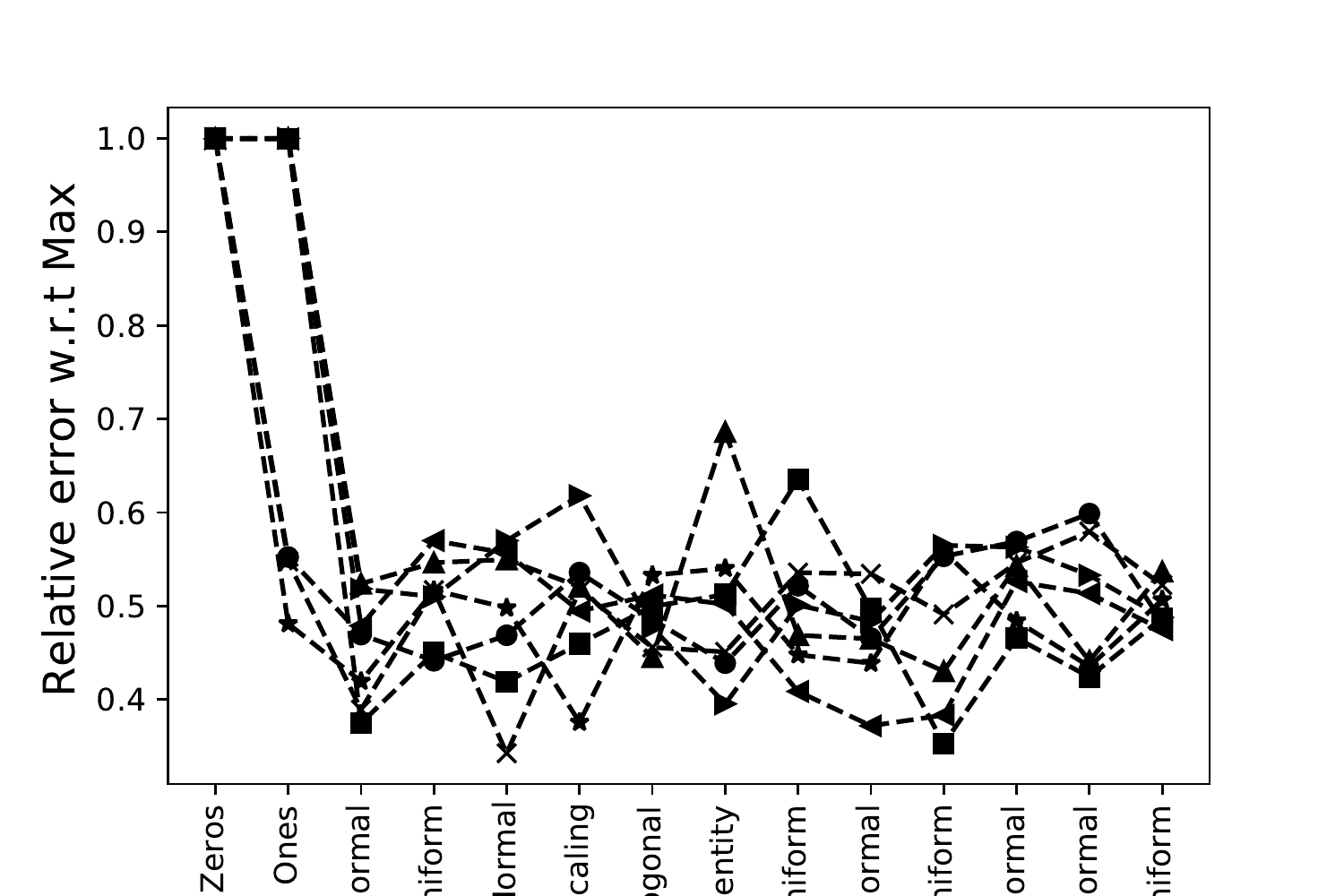}
		
		\caption{Same as Fig.~\ref{fig:Teff_tests} but for \logg .}
		\label{fig:Logg_test}
	\end{figure*}

	The accuracies for gravity behave differently than the one of \Teff\ with respect to the various parameters. According to Fig.~\ref{fig:Logg_test}, the optimal values are found to be relu or $\tanh$ for the Activation function; Adam, Adamax or RMSprop for the optimizer, a number of Batches between 32 and 128, an epoch number of 3000, a Dropout fraction between 0.3 and 0.4, a mean squared logarithmic error loss function, and all kinds of initialisers except for Zeros and Ones.
	
	In case of \logg\, the optimal configuration is found to be using the following parameters:
	\begin{itemize}
		\item[] Activation function: $\tanh$.
		\item[] Optimizer: Adamax.
		\item[] Batches: 128.
		\item[] Epochs: 3000.
		\item[] Dropout: 30\%.
		\item[] Loss function: mean squared logarithmic error. 
		\item[] Kernel initialiser: he\_normal. 
	\end{itemize}

	\subsection{Metallicity} 
	\label{meta}
\begin{figure*}[!h]
		\centering
		\includegraphics[scale=0.35]{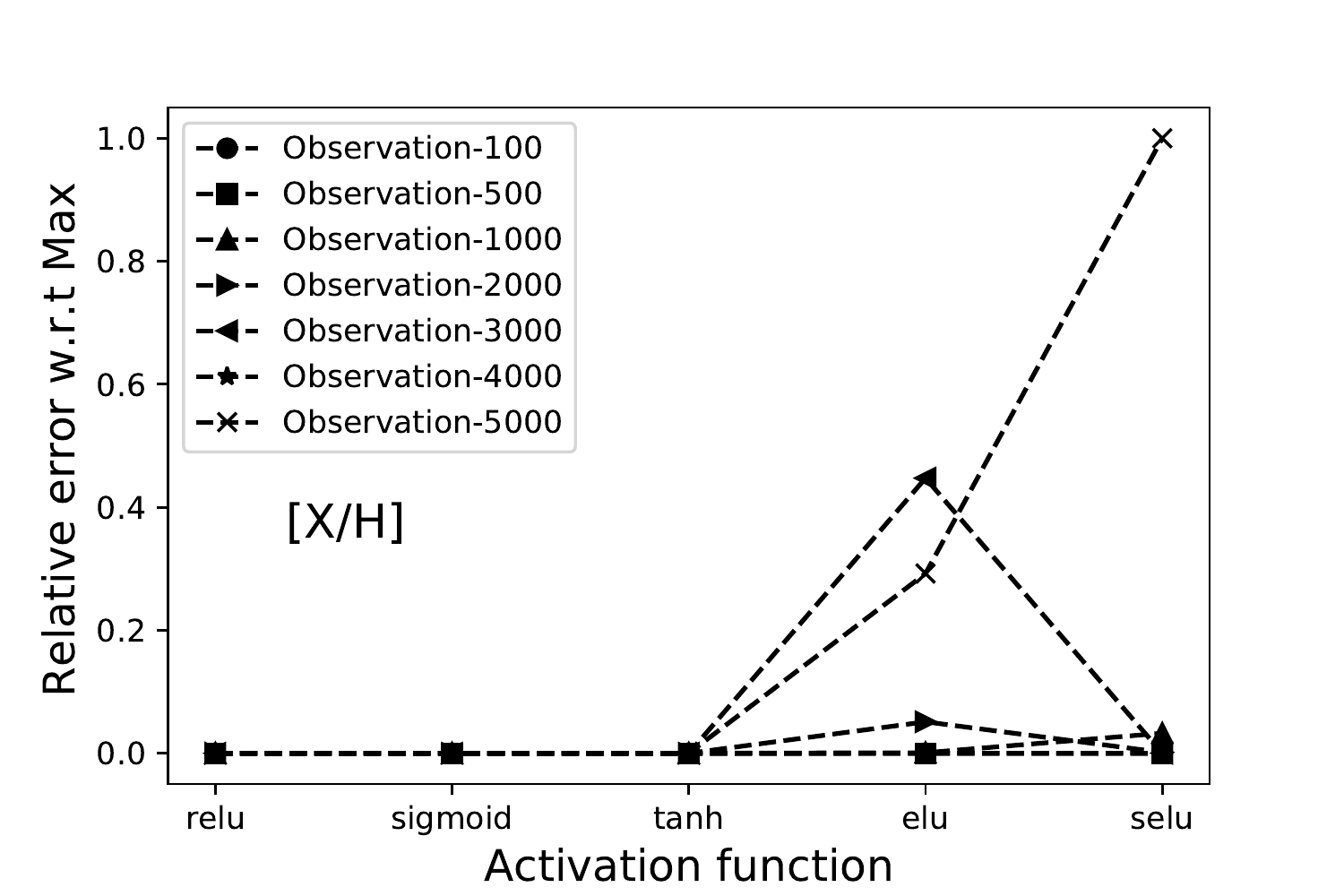}
		\includegraphics[scale=0.35]{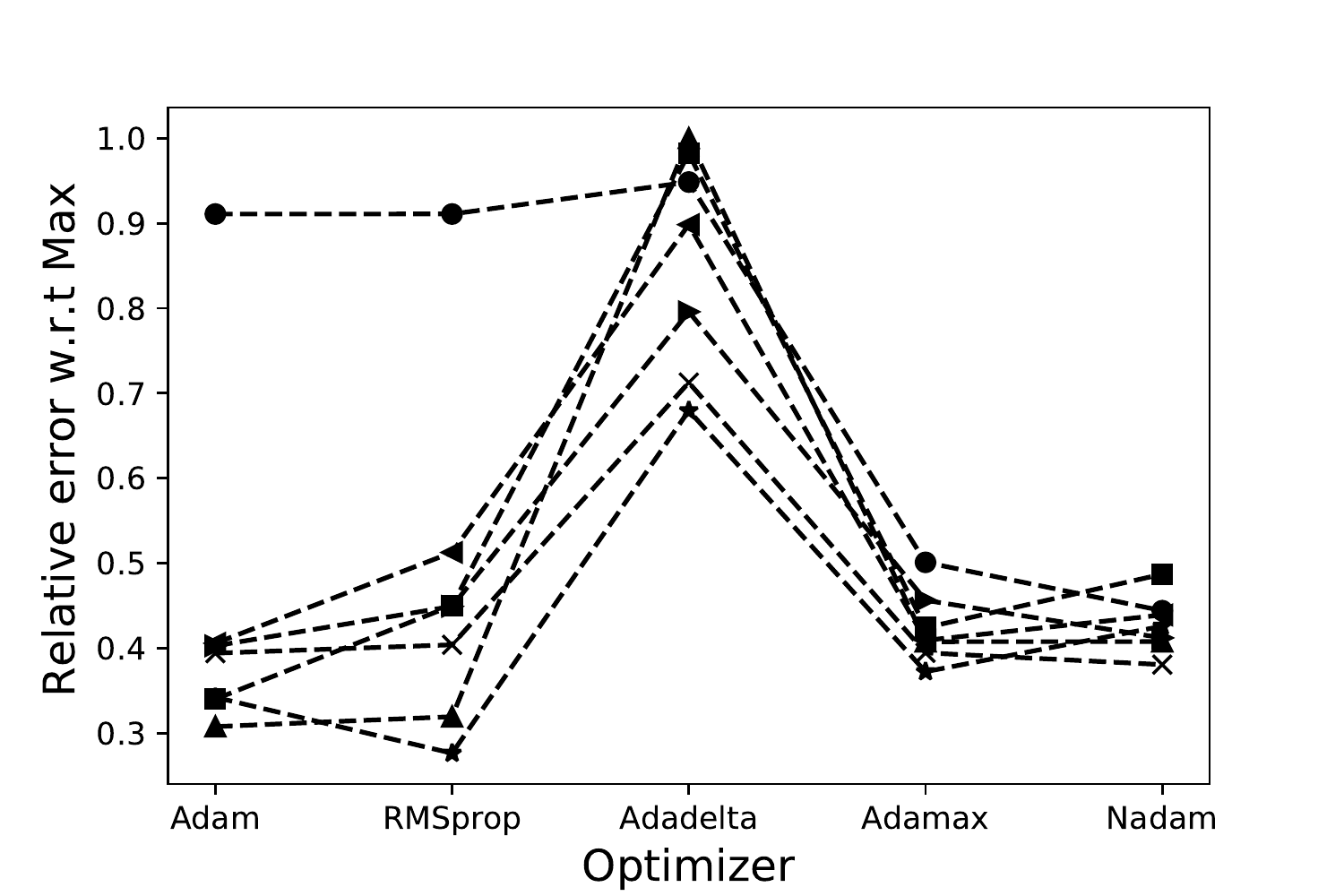}
		\includegraphics[scale=0.35]{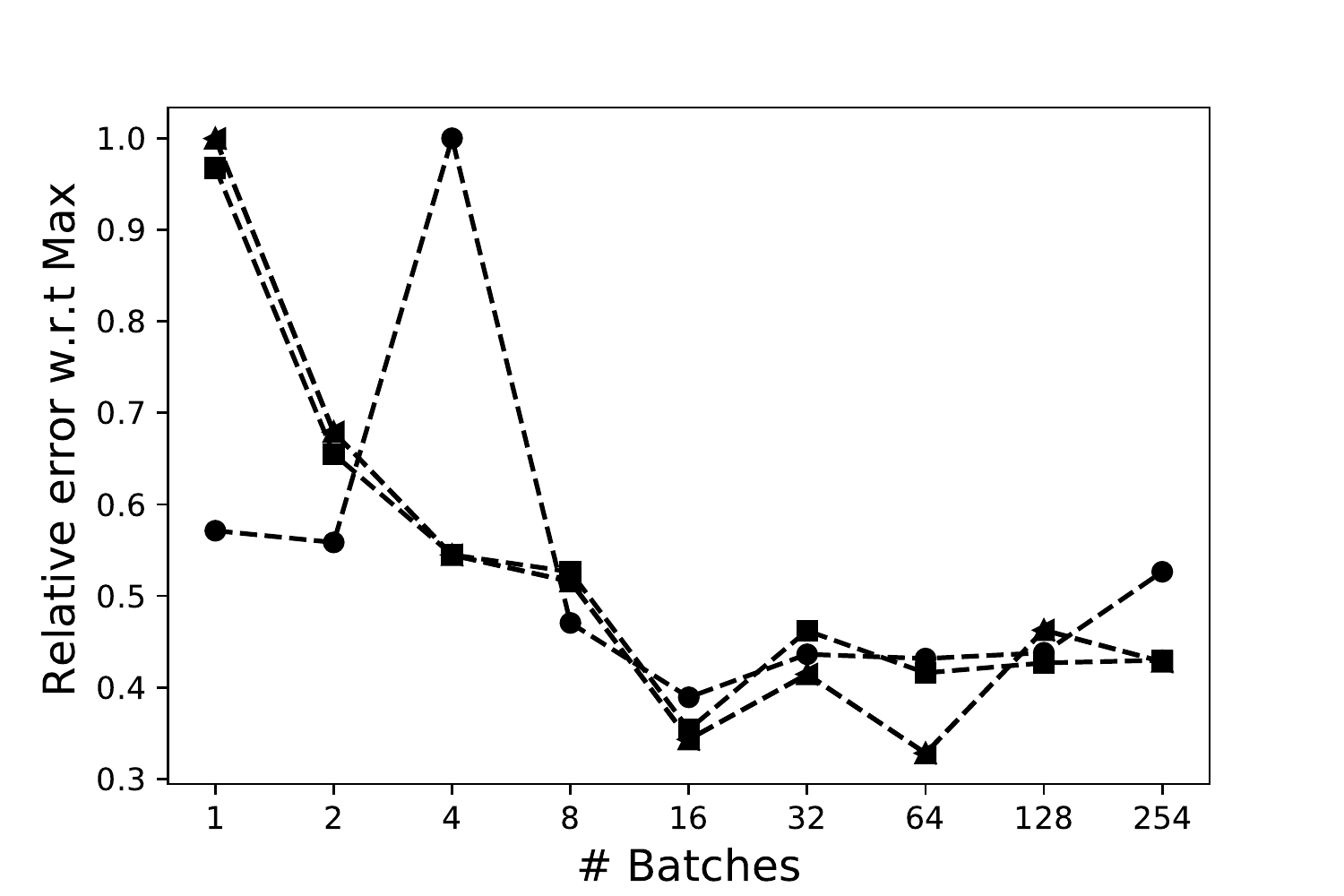}\\
		\includegraphics[scale=0.35]{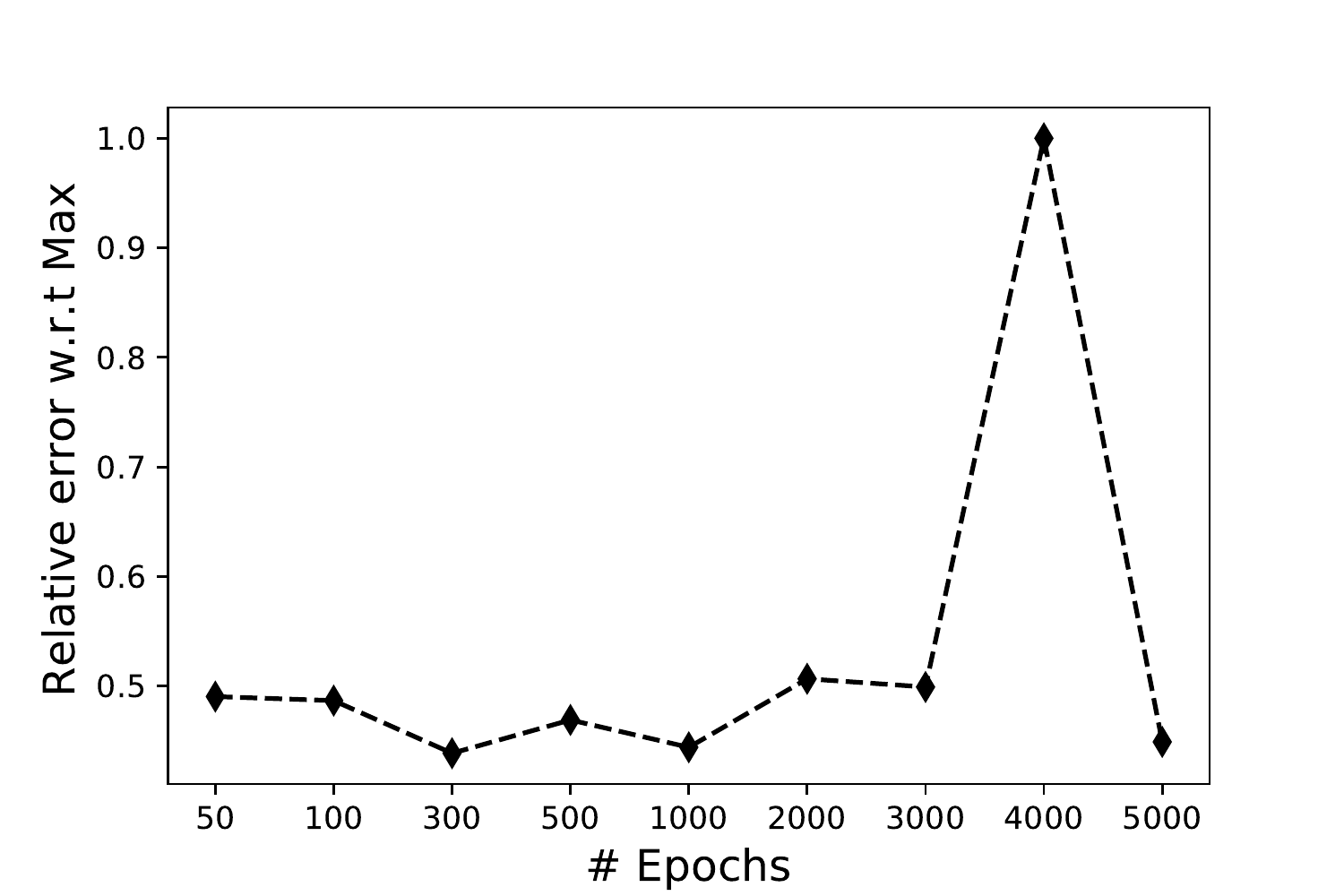}
		\includegraphics[scale=0.35]{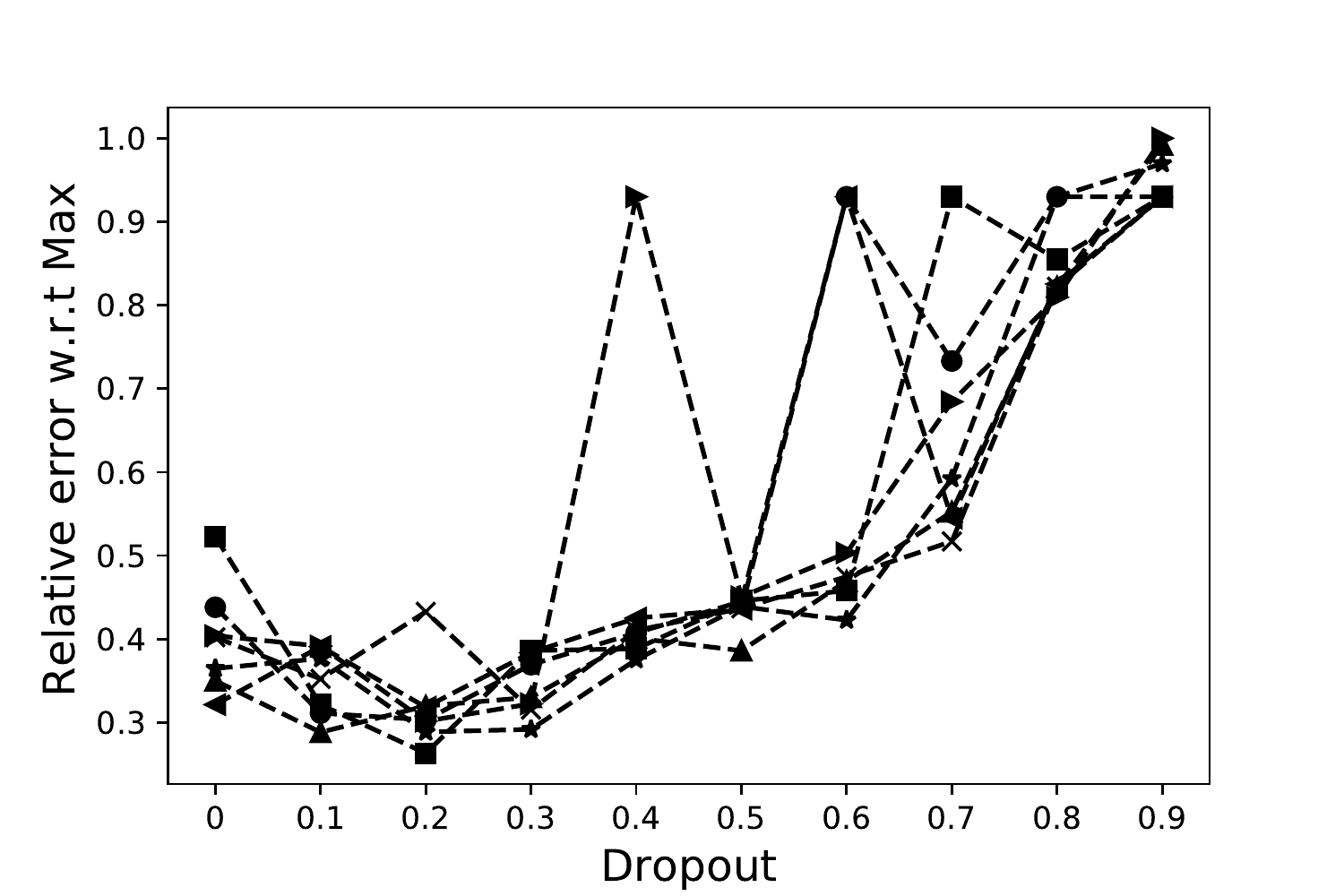}
		\includegraphics[scale=0.35]{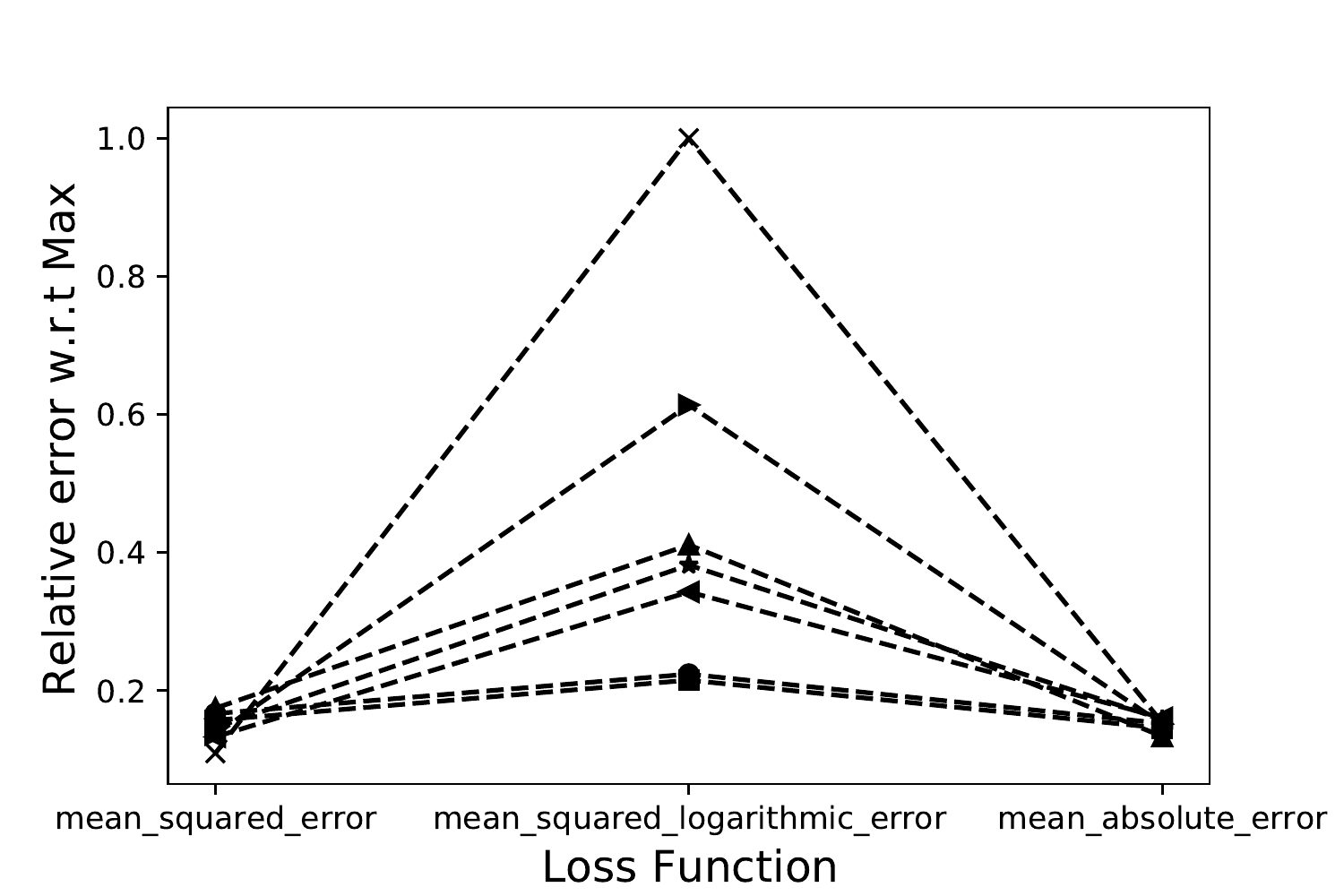}\\
		\includegraphics[scale=0.35]{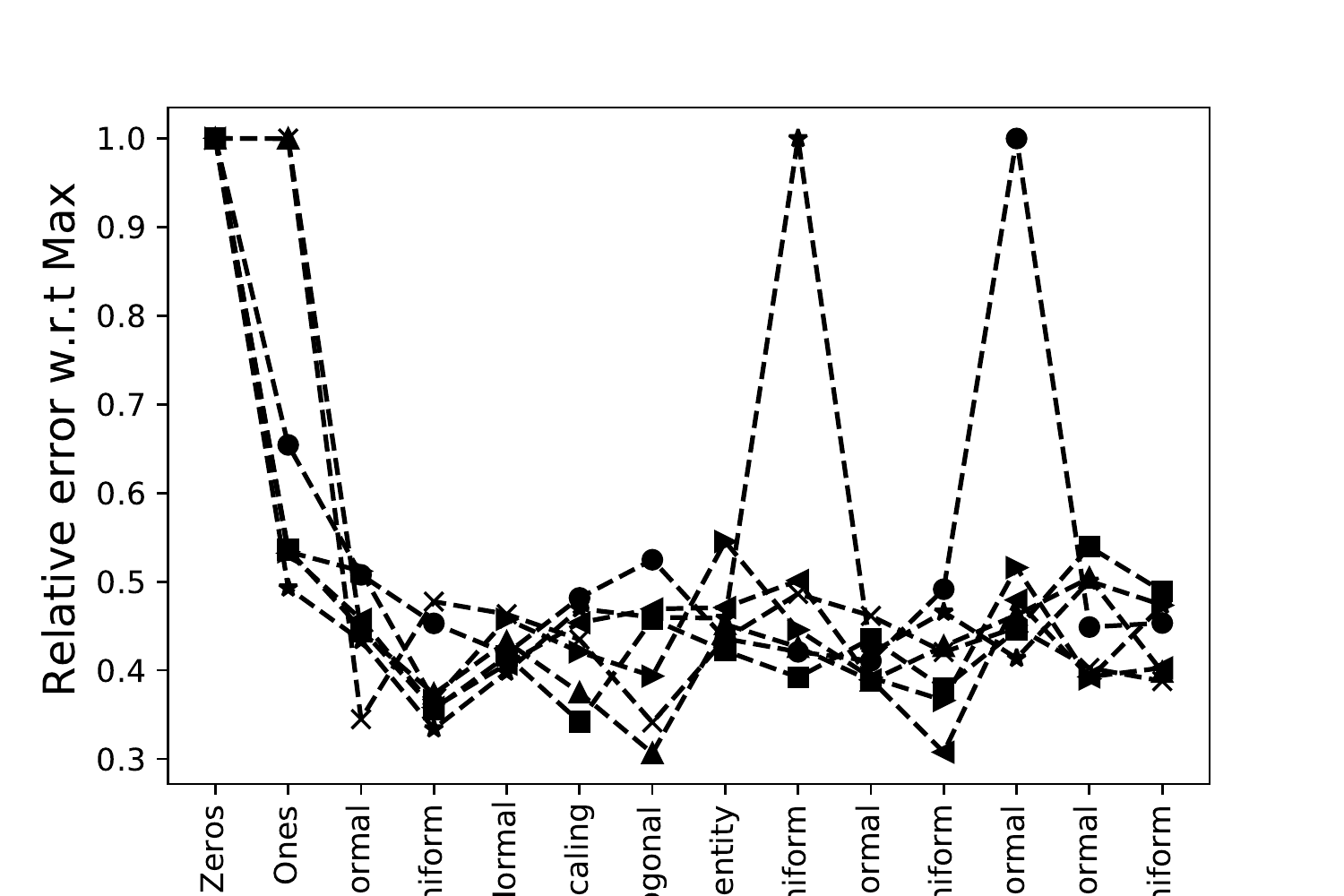}
		
		\caption{Same as Fig.~\ref{fig:Teff_tests} but for \met .}
		\label{fig:Meta_test}
	\end{figure*}

	The metallicity parameter, \met, also behaves differently than \Teff\ and \logg. As seen in Fig.~\ref{fig:Meta_test}, \met\ requires a different combination of parameters in our CNN in order to reach optimal results. $\tanh$ or relu activation functions give the least error in most epochs number situations. Adam and RMSprop optimizer lead to similar results within few percents of differences. A combination of 16 Batches and 1000 epochs is appropriate to derive \met\ with low errors. A dropout between 10\% and 30\%, a mean absolute error for a loss function and a RandomUniform kernel initialiser are to be used in order to reach the highest possible accuracy for \met. Our technique was applied to A stars and extrapolated to FGK stars (Sec.~\ref{fgk}). However, specific considerations should be taken into account when deriving the metallicities of cool stars due to forests of molecular lines that are present in the spectra (\cite{2021arXiv211114950P}).

	In case of \met, the optimal configuration is found to be using the following parameters:
	\begin{itemize}
		\item[] Activation function: $\tanh$.
		\item[] Optimizer: Adam.
		\item[] Batches: 16.
		\item[] Epochs: 1000.
		\item[] Dropout: 20\%.
		\item[] Loss function: mean absolute error. 
		\item[] Kernel initialiser: RandomUniform. 
	\end{itemize}

	\subsection{Projected Equatorial Rotational Velocity} 
	\label{vsini}
	
\begin{figure*}[!h]
		\centering
		\includegraphics[scale=0.35]{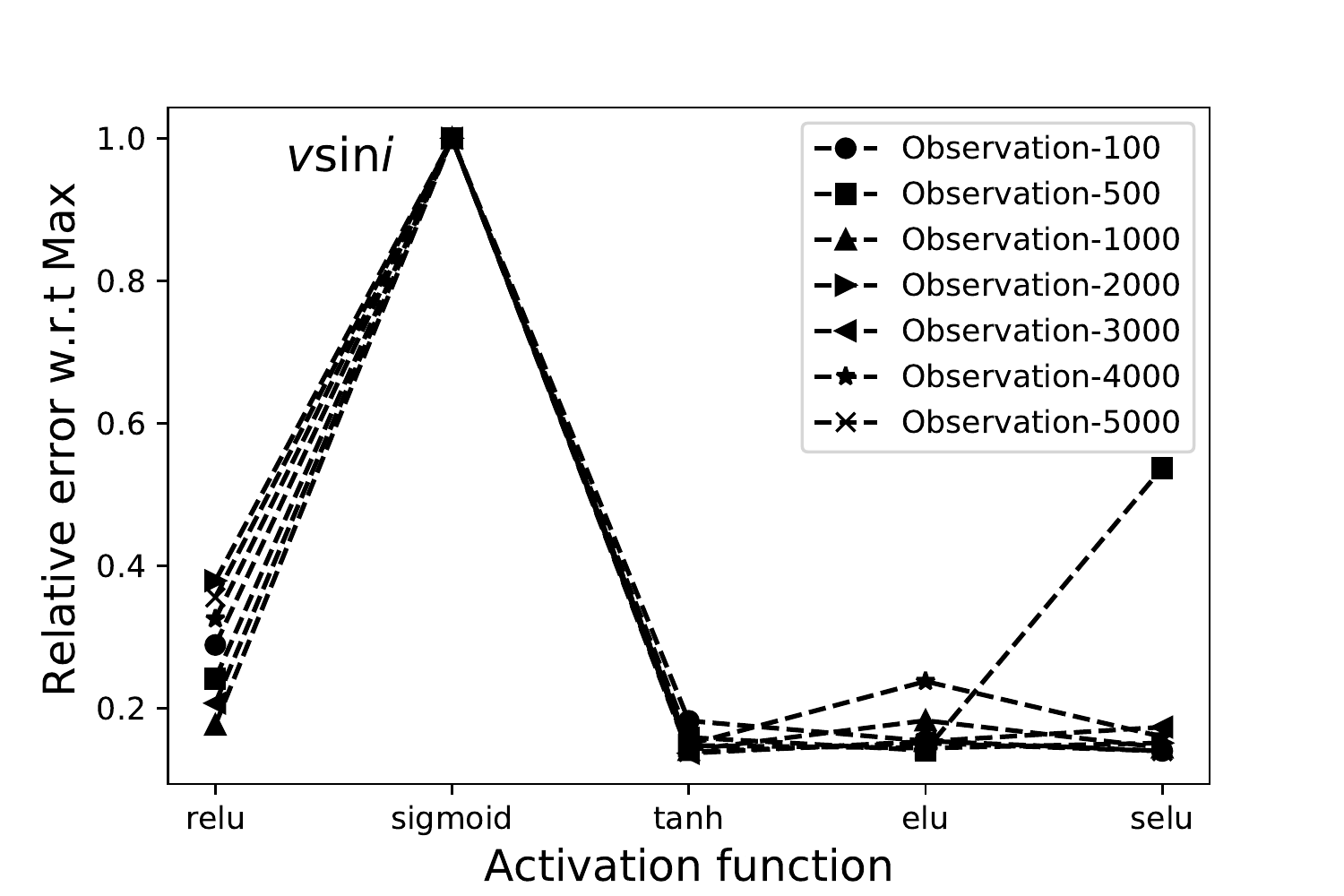}
		\includegraphics[scale=0.35]{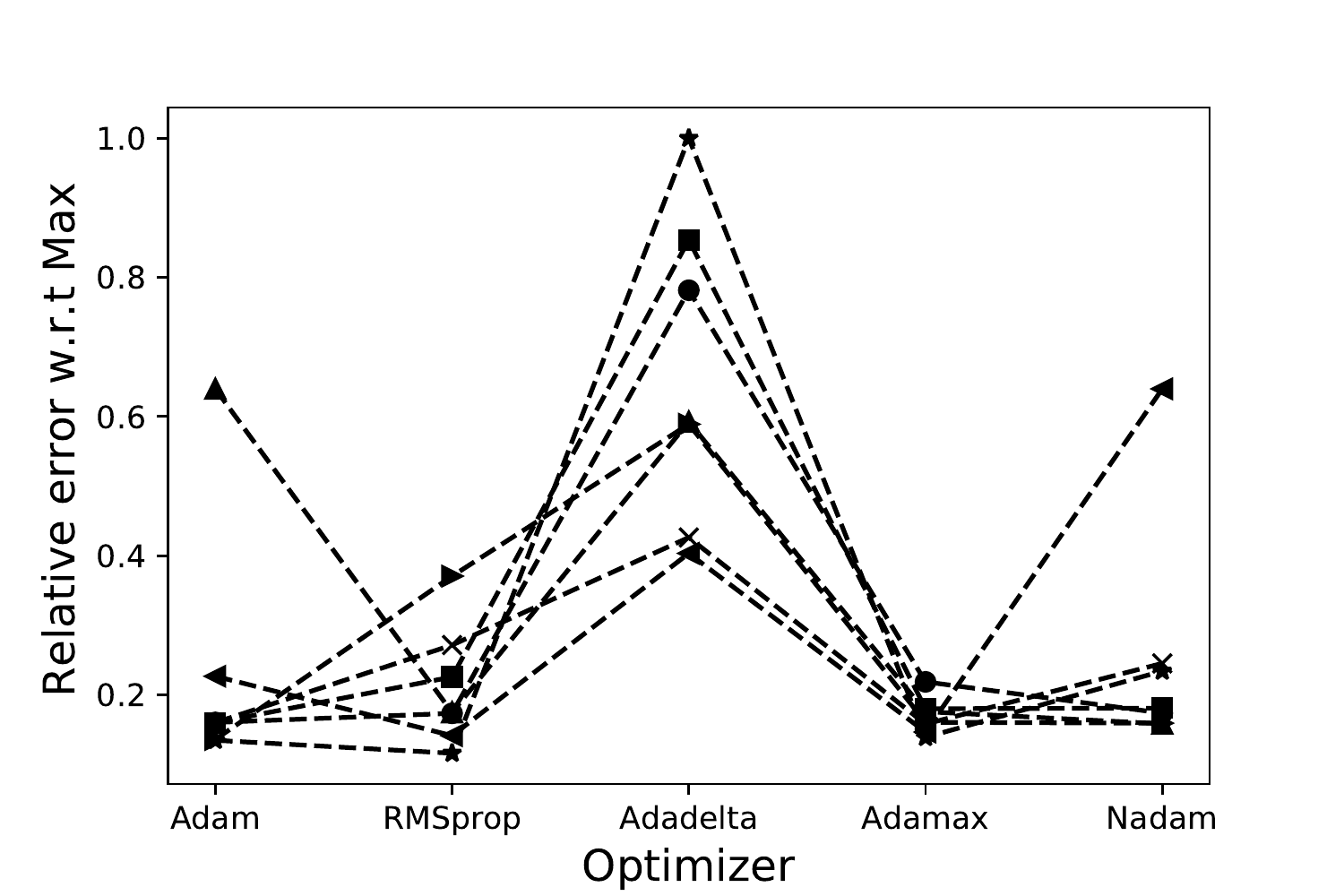}
		\includegraphics[scale=0.35]{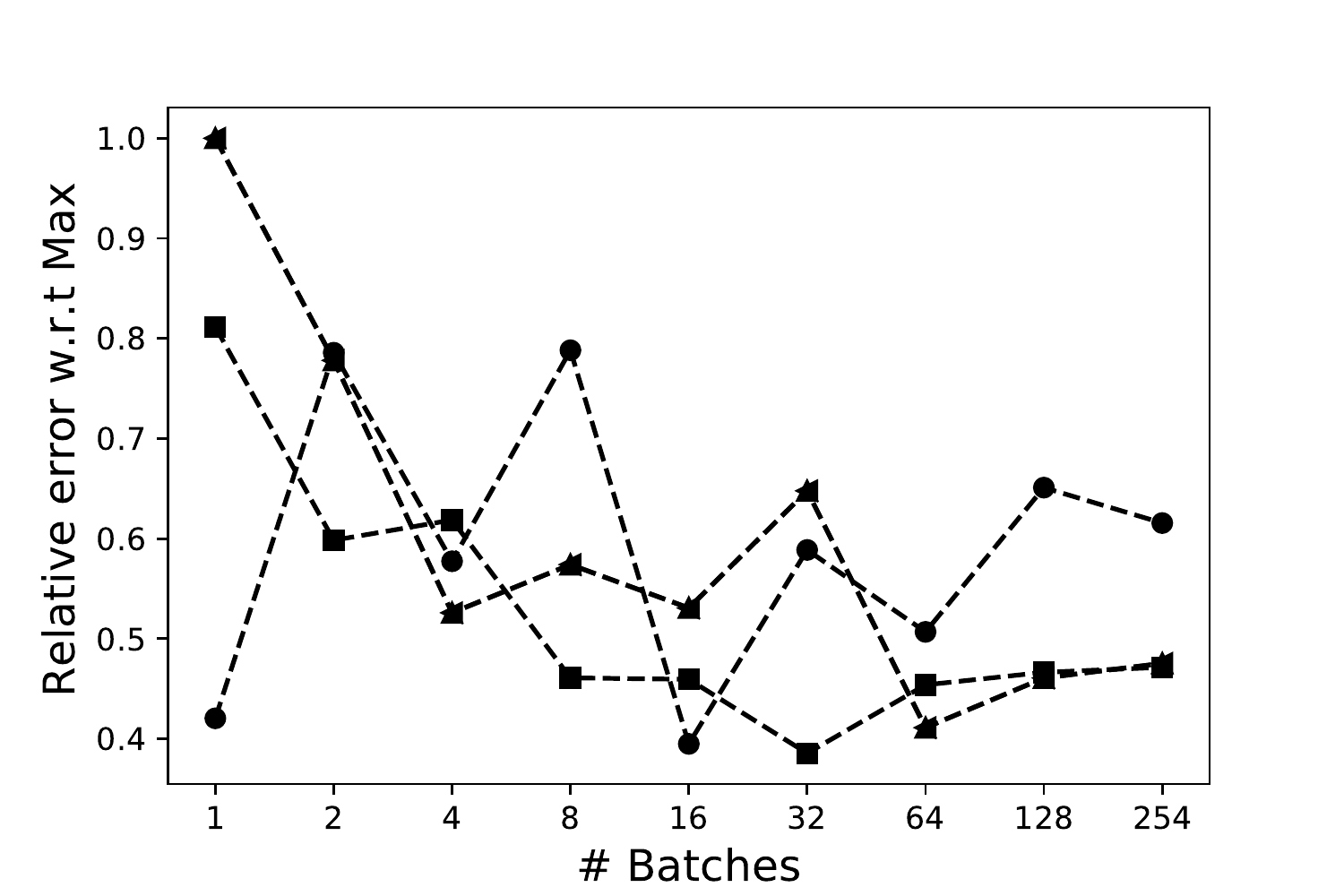}\\
		\includegraphics[scale=0.35]{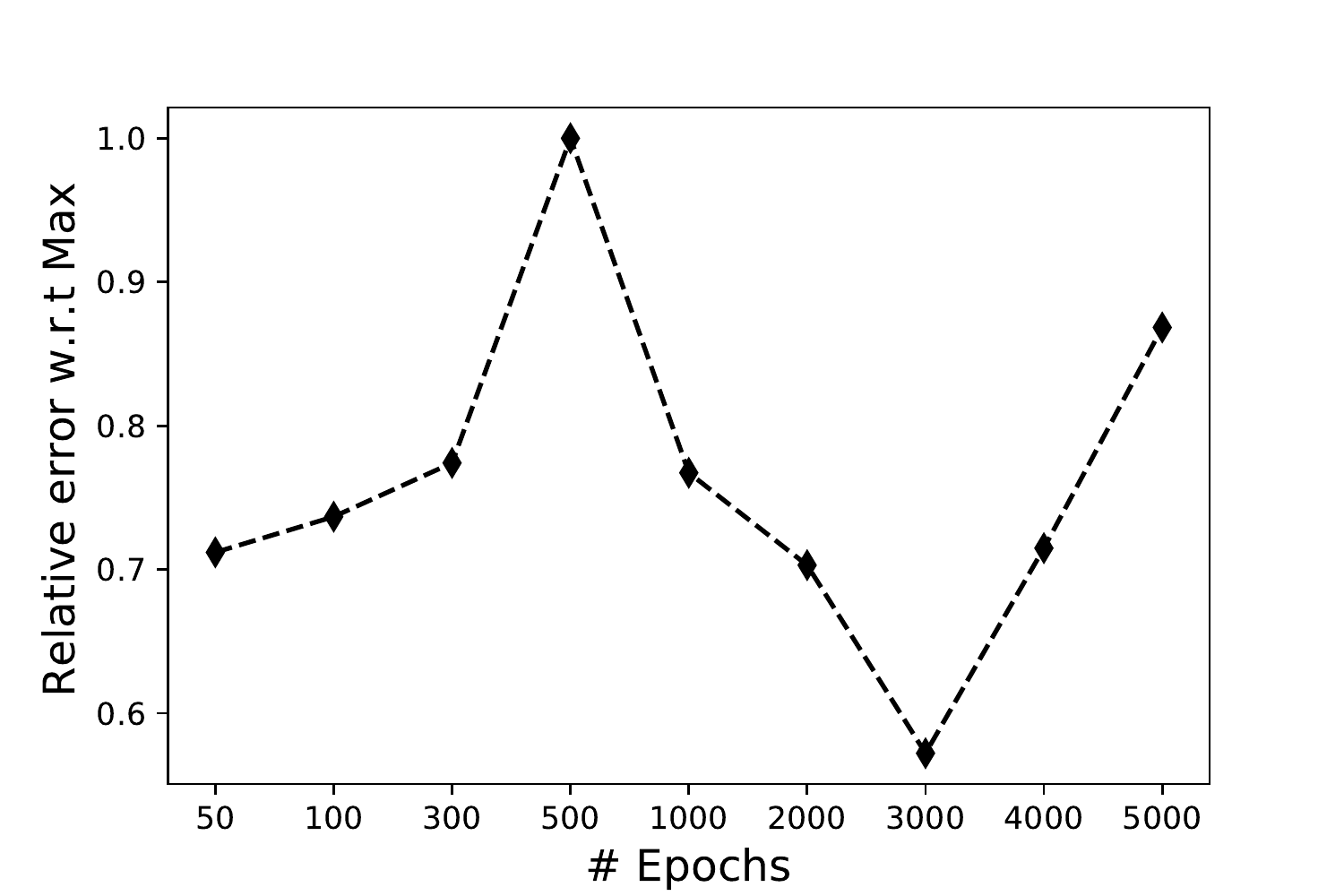}
		\includegraphics[scale=0.35]{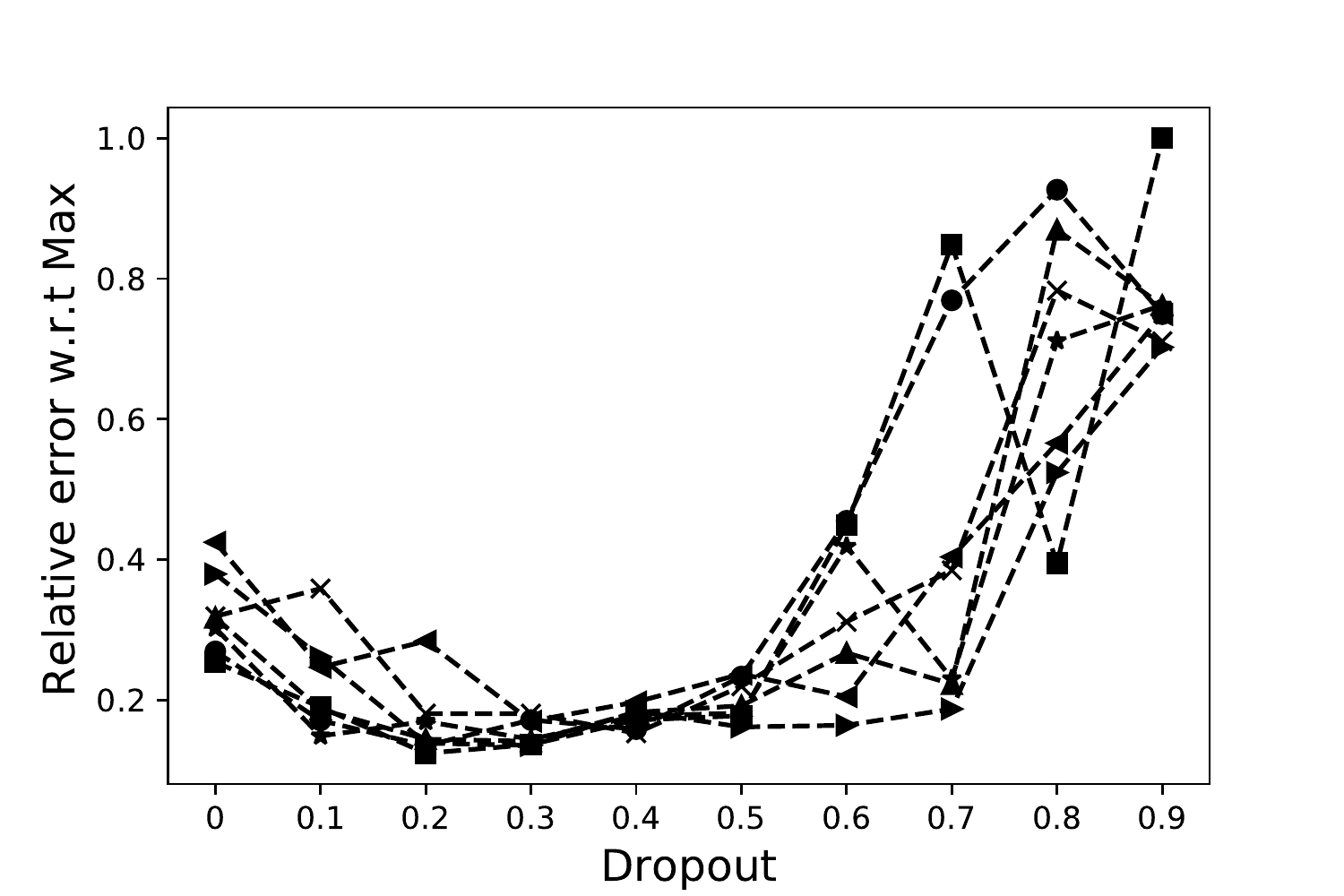}
		\includegraphics[scale=0.35]{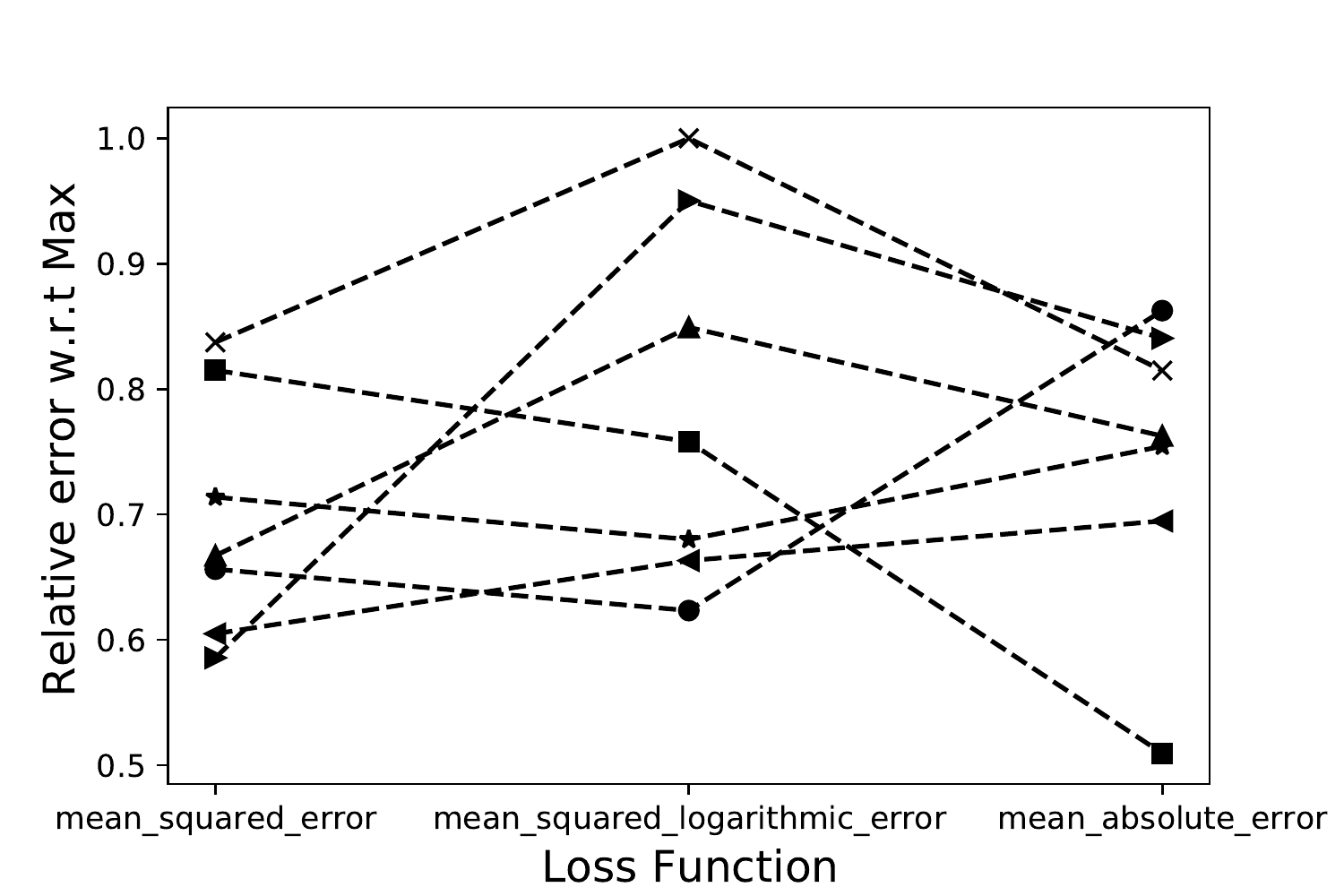}\\
		\includegraphics[scale=0.35]{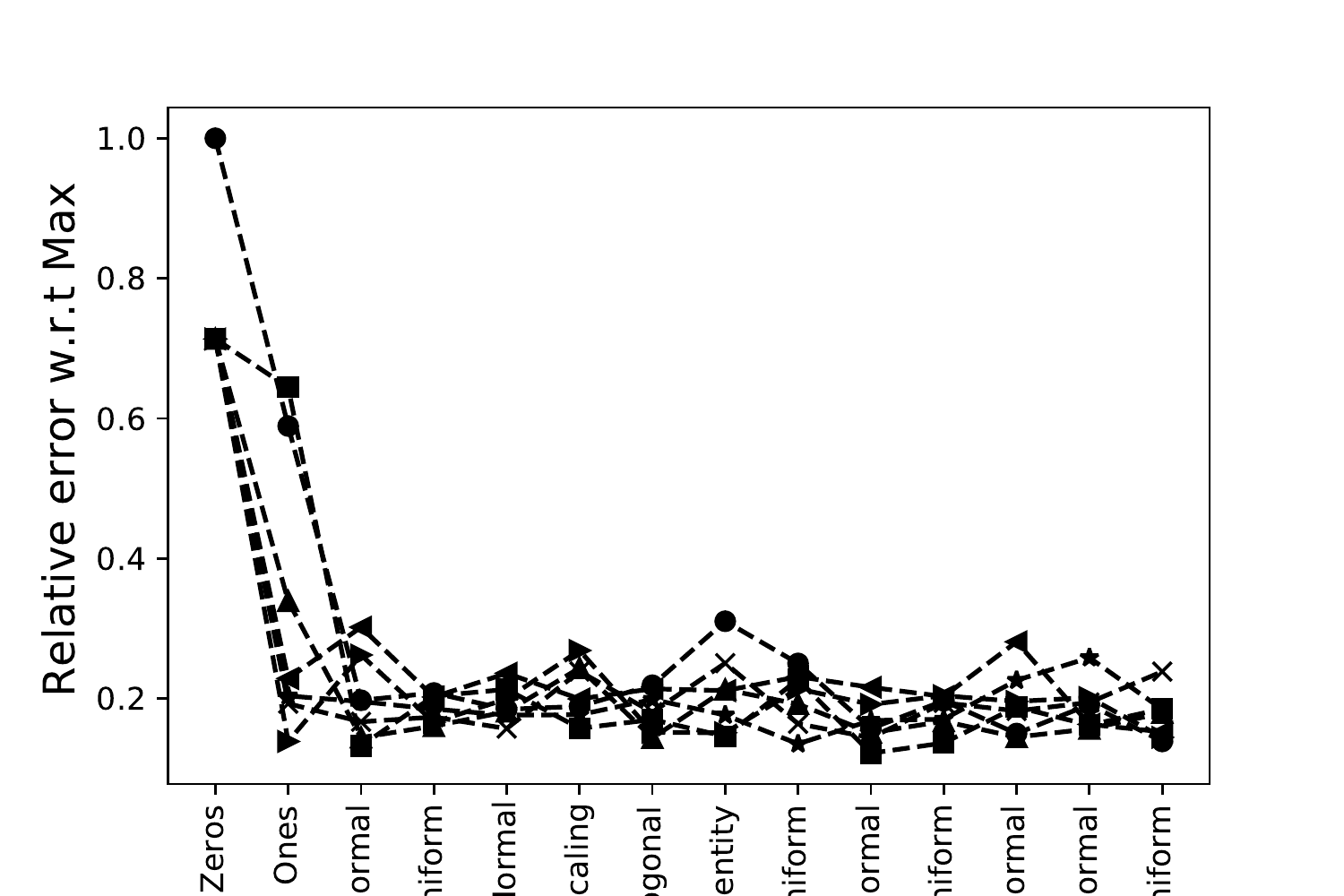}
		
		\caption{Same as Fig.~\ref{fig:Teff_tests} but for \vsini .}
		\label{fig:Vrot_test}
	\end{figure*}

	Finally, in case of the equatorial projected rotational velocity, \vsini, $\tanh$ seems to be the optimal activation function independently of the epochs and Batches number (Fig.~\ref{fig:Vrot_test}). Adam or Adamax optimizers can be used for \vsini\ with small differences in the derived accuracies. A combination of 32 Batches with 3000 epochs is the one that gives the minimum error for the derived \vsini\ values. A dropout fraction between 0.1 and 0.4 yields very close errors. A mean squared error can be used for the loss function and all kernel initialisers can also be applied except the Zeros and Ones for the same reason explained in Sec.~\ref{teff}.

	In case of \vsini, the optimal configuration is found to be using the following parameters:
	\begin{itemize}
		\item[] Activation function: $\tanh$.
		\item[] Optimizer: Adamax.
		\item[] Batches: 32.
		\item[] Epochs: 3000.
		\item[] Dropout: 30\%.
		\item[] Loss function: mean squared error. 
		\item[] Kernel initialiser: he\_Uniform. 
	\end{itemize}

	\subsection{Database Size and the role of Augmentation}  
	\label{small-big}
	In order to check the dependency of the performance of the CNN on the size of the training set, three databases are used. The first database (TDB1) contains 25\,000 random synthetic spectra as explained in Sec.~\ref{results} ,  the second database (TDB2) contains 40\,000 random spectra, and the third (TDB3) contains 70\,000 spectra, resulting from the same TDB1 parameter ranges. We have also checked the importance of using Data Augmentation as a regularization technique for deriving accurate parameters (see Sec.~\ref{data_aug} for details). \\
	
\begin{figure}[!h]
		\centering
		\includegraphics[scale=0.4]{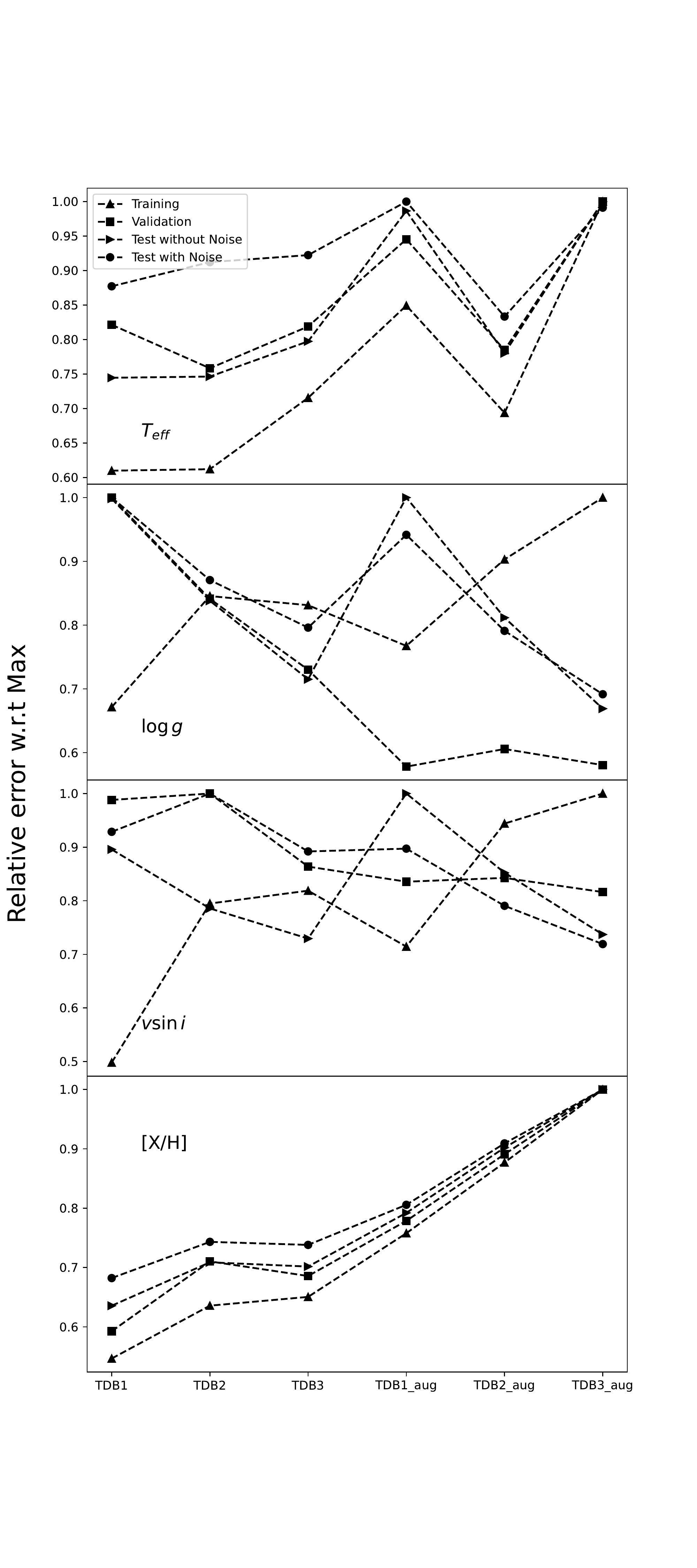}
		\caption{Relative errors for each stellar parameter using TDB1, TDB2 and TDB3 with and without data augmentation as training dataset.}
		\label{fig:aug}
	\end{figure}

	For each stellar parameter, we used the optimal CNN with the configuration that was derived in Secs.~\ref{teff} to \ref{vsini}. Each configuration was tested with TDB1, TDB2 and TDB3 with and without Data Augmentation. Fig.~\ref{fig:aug} displays the average relative standard deviation for each stellar parameter with the respect to the maximum values, for the training, validation, test and observation sets. In order to quantify these proxies for the uncertainties of the techniques, Tab.~\ref{Tab:aug} collects the standard deviations for the 4 stellar parameters as a function of the training database. 
	
	According to Tab.~\ref{Tab:aug}, each parameter behaves differently with respect to the change of the databases. This is mainly due to the number of unique values of the parameter in the database. 
	For that reason, \met\ is well represented by TDB1 without data augmentation whereas \Teff\ \logg\, and \vsini\ require a larger database to be well represented. \logg\ can be well represented with TDB3 with data augmentation whereas \Teff\ can be predicted with TDB2 with data augmentation. Finally, \vsini\ is can be predicted using TDB3 with data augmentation.

	\begin{table*}[!h]
		\centering
		\begin{tabular}{||c|c|c|c|c||}
			\hline \hline
			Database&\multicolumn{4}{c||}{TDB1} \\ \cline{2-5}
			&  $\sigma_{\Teff}$ (K)& $\sigma_{\logg}$ (dex)& $\sigma_{\met}$ (dex)& $\sigma_{\vsini}$ \kms \\
			\hline
			Training&  78  &0.03 &0.06&0.97 \\
			Validation&112  & 0.11&0.07 &4.00\\
			Test without Noise&   98&0.10&0.07  &3.30     \\
			Test with Noise&  133&0.13 &0.07&5.25    \\
			\hline
			&\multicolumn{4}{c||}{TDB1 with Data Augmentation}\\ \cline{2-5}
			&  $\sigma_{\Teff}$ (K)& $\sigma_{\logg}$ (dex)& $\sigma_{\met}$ (dex)& $\sigma_{\vsini}$ \kms \\
			\hline
			Training& 109&0.03&0.09&1.40         \\
			Validation&129&0.07&0.09&3.36  \\
			Test without Noise&129&0.10&0.10&3.66              \\
			Test with Noise& 152&0.12&0.10&5.00     \\
			\hline
			&\multicolumn{4}{c||}{TDB2}\\ \cline{2-5}
			&  $\sigma_{\Teff}$ (K)& $\sigma_{\logg}$ (dex)& $\sigma_{\met}$ (dex)& $\sigma_{\vsini}$ \kms \\
			\hline
			Training& 79&0.04&0.08&1.55         \\
			Validation&104&0.10&0.08&4.00        \\
			Test without Noise&99&0.09&0.09&2.90              \\
			Test with Noise& 139&0.11&0.09&5.50      \\
			\hline
			&\multicolumn{4}{c||}{TDB2 with Data Augmentation}\\ \cline{2-5}
			&  $\sigma_{\Teff}$ (K)& $\sigma_{\logg}$ (dex)& $\sigma_{\met}$ (dex)& $\sigma_{\vsini}$ \kms \\
			\hline
			Training& 89&0.04&0.10&1.85         \\
			Validation&107&0.07&0.10&3.40            \\
			Test without Noise&103&0.09&0.10&3.12              \\
			Test with Noise&127&0.10&0.11&4.36       \\
			\hline
			&\multicolumn{4}{c||}{TDB3}\\ \cline{2-5}
			&  $\sigma_{\Teff}$ (K)& $\sigma_{\logg}$ (dex)& $\sigma_{\met}$ (dex)& $\sigma_{\vsini}$ \kms \\
			\hline
			Training& 92&0.04& 0.07& 1.60      \\
			Validation& 112&0.08&0.08&   3.50  \\
			Test without Noise&105&0.08&0.08&2.70             \\
			Test with Noise&  140& 0.10&0.09& 4.90 \\
			\hline
			&\multicolumn{4}{c||}{TDB3 with Data Augmentation}\\ \cline{2-5}
			&  $\sigma_{\Teff}$ (K)& $\sigma_{\logg}$ (dex)& $\sigma_{\met}$ (dex)& $\sigma_{\vsini}$ \kms \\
			\hline
			Training& 128&0.04&0.11& 1.95\\
			Validation& 136&   0.06& 0.11&3.20       \\
			Test without Noise&131& 0.07&  0.11&2.70          \\
			Test with Noise& 150& 0.08&0.12& 3.90  \\
			\hline
		\end{tabular}
		\caption{Derived standard deviation for each parameters using TDB1, TDB2 and TDB3 with and without data augmentation. The values for the Training, Validation, and the two sets of Test are depicted in this table.}
		\label{Tab:aug}
	\end{table*}

	\subsection{Accuracy for the Optimal Configuration}
\begin{figure*}[!h]
		\centering
		\includegraphics[scale=0.5]{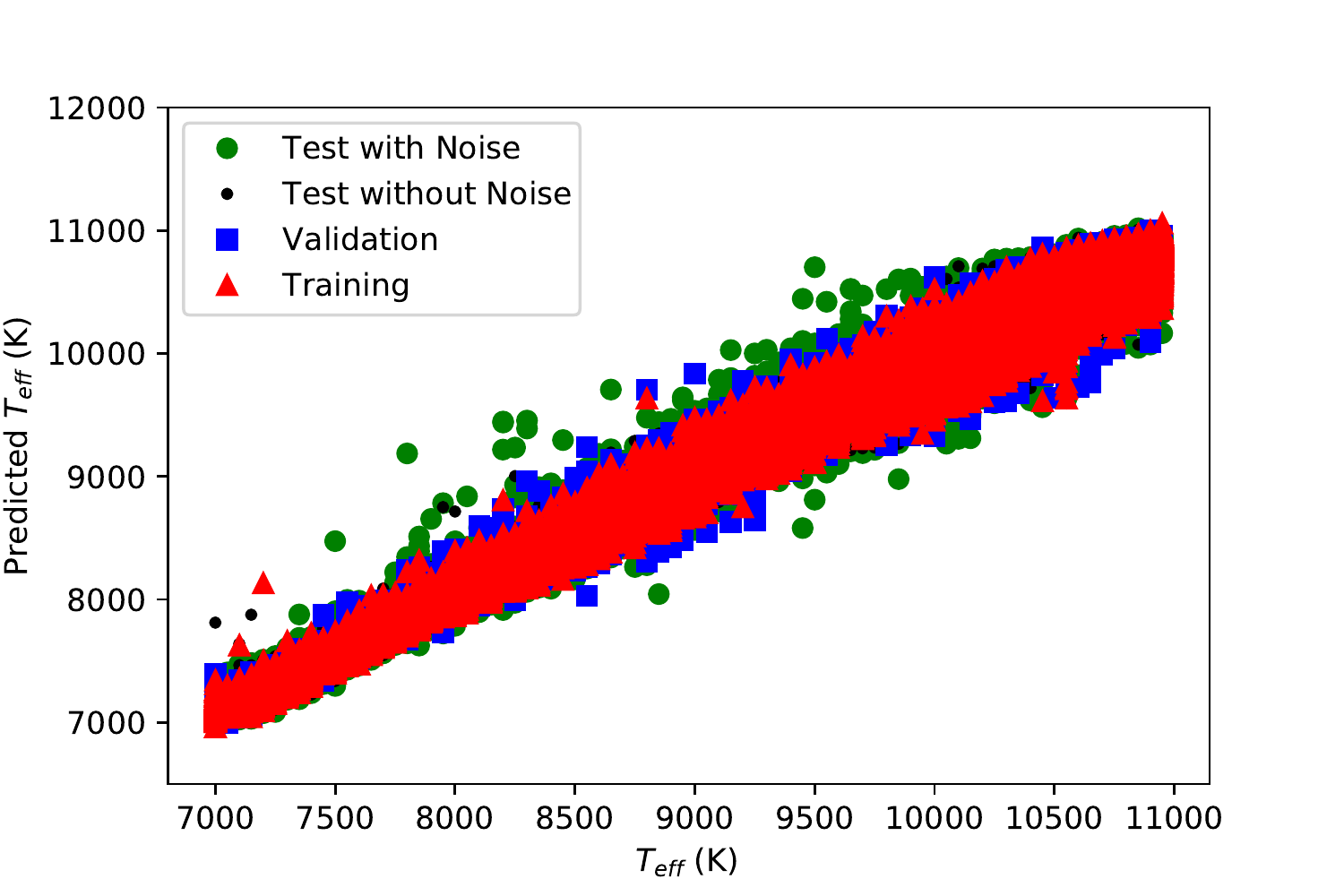}
		\includegraphics[scale=0.5]{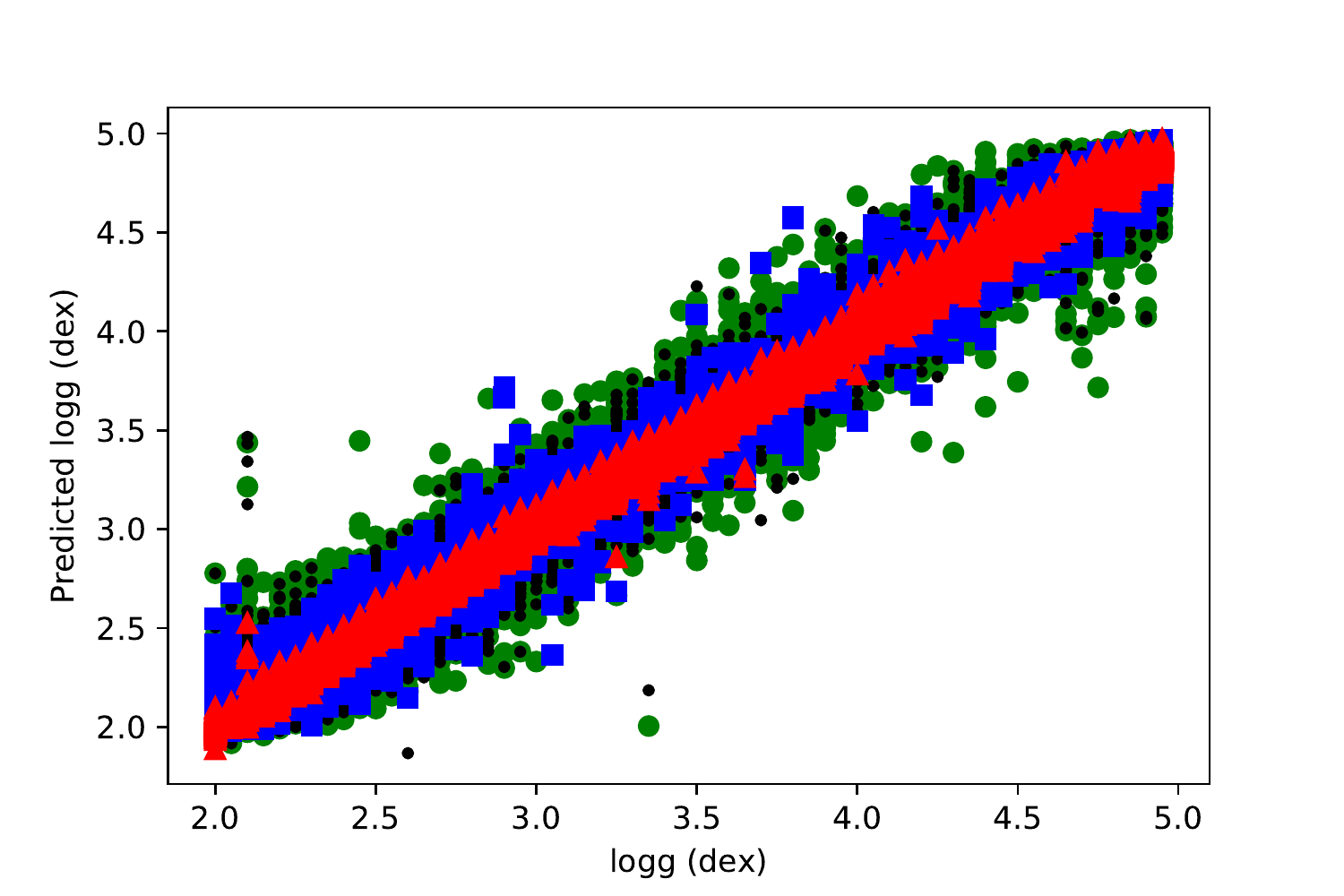}\\
		\includegraphics[scale=0.5]{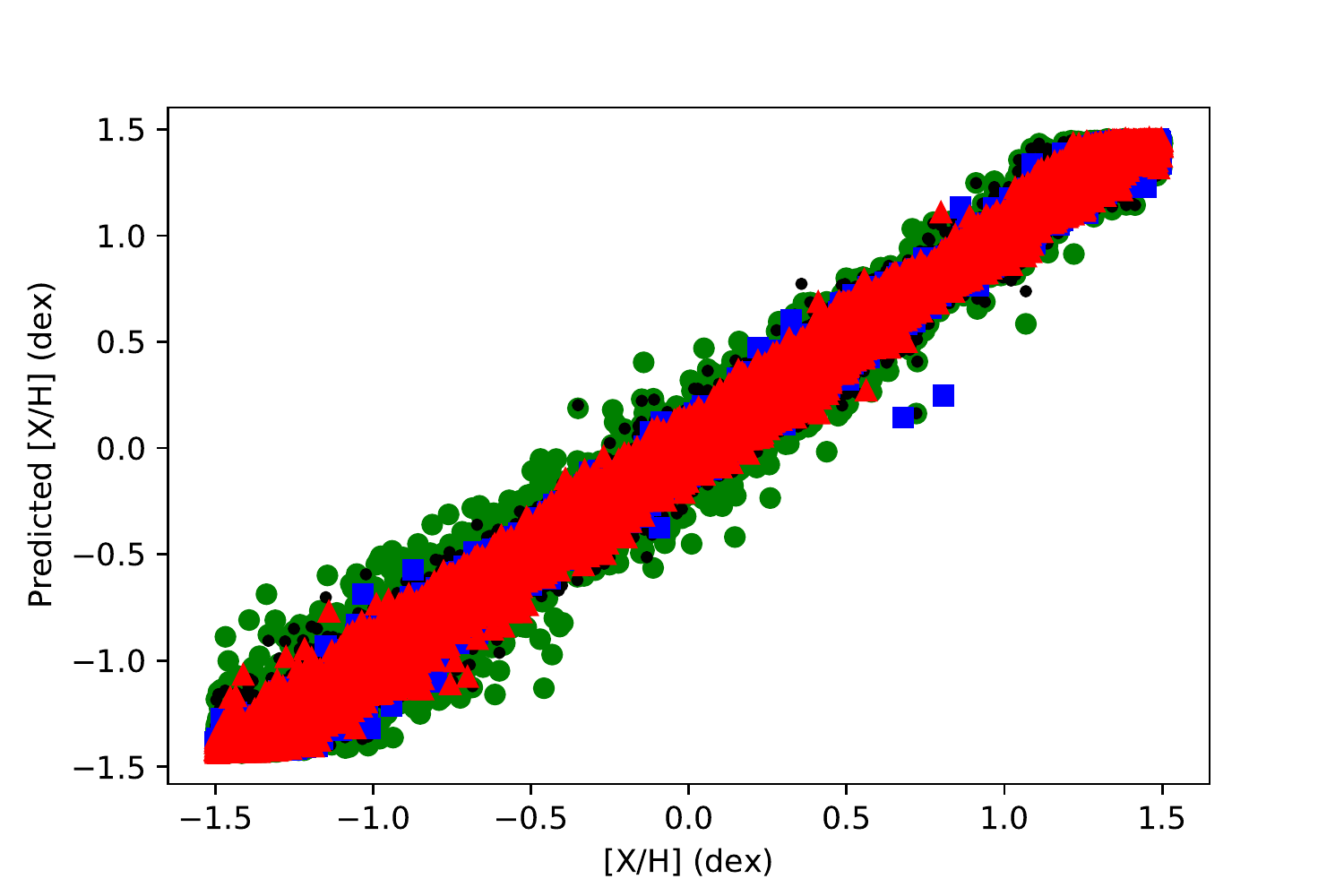}
		\includegraphics[scale=0.5]{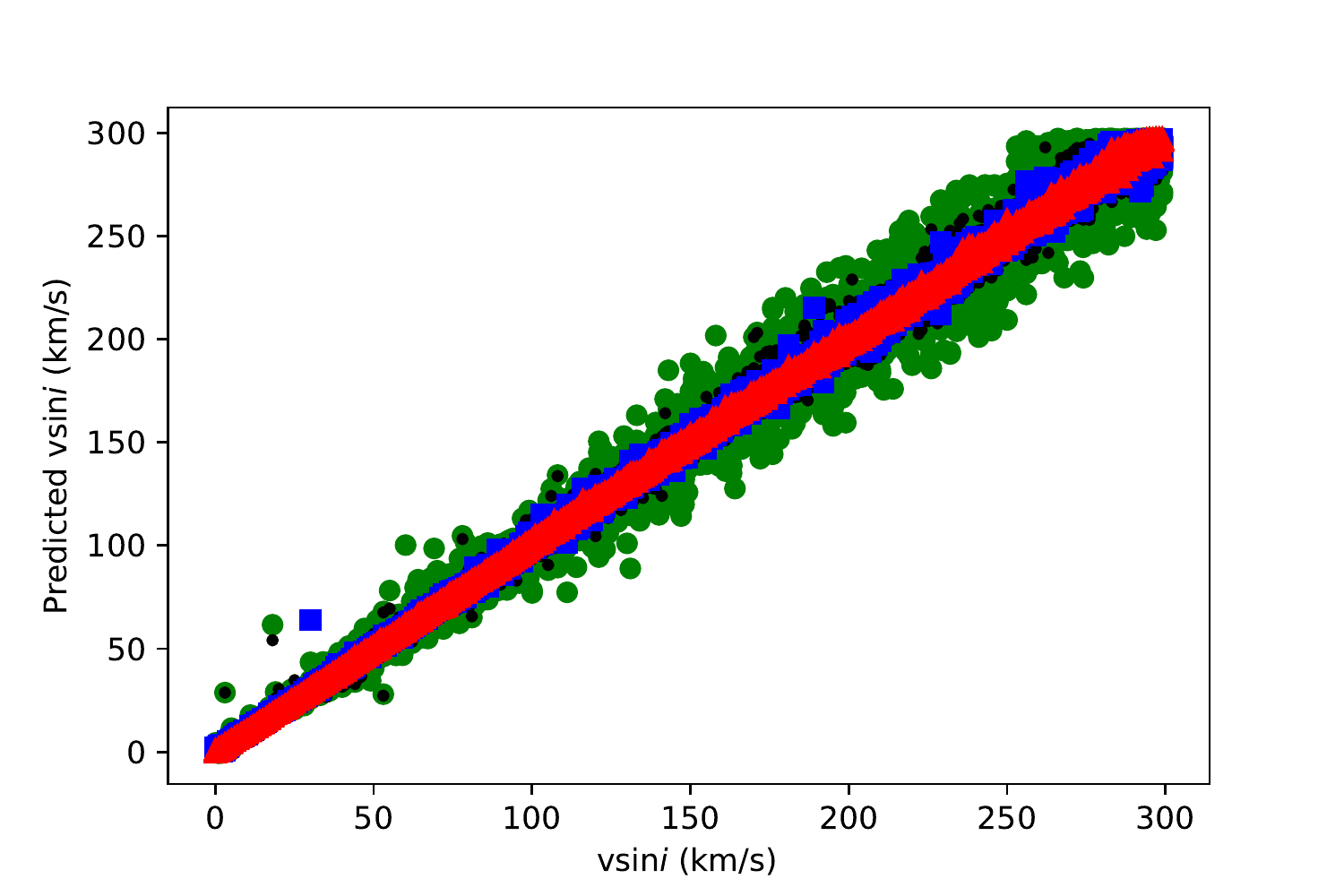}
		\caption{Predicted stellar parameters using the optimal CNN configurations for \Teff, \logg, \vsini, and \met\ as a function of the input ones for the training, validation and test databases as well as for the noise added observations. }
		\label{fig:optimal_values}
	\end{figure*}

	After selecting the optimal configuration for each stellar parameter, the predicted parameters are displayed in Fig.~\ref{fig:optimal_values} as a function of the input ones for the training, validation, and the two sets of Test datasets. All data points are located around the $y=x$ line. The dispersion of the observation around that line is due to spectra with very low signal to noise.  The accuracy that we found using our CNN architecture seem to be appropriate for A stars as they are comparable to most of previous studies using classical tools (\cite{2014sf2a.conf..451A}) or more complicated statistical tools (\cite{Gebran, 2019OAst...28...68K}). The same is true for all parameters. 
	
\begin{figure*}[!h]
		\centering
		\includegraphics[scale=0.45]{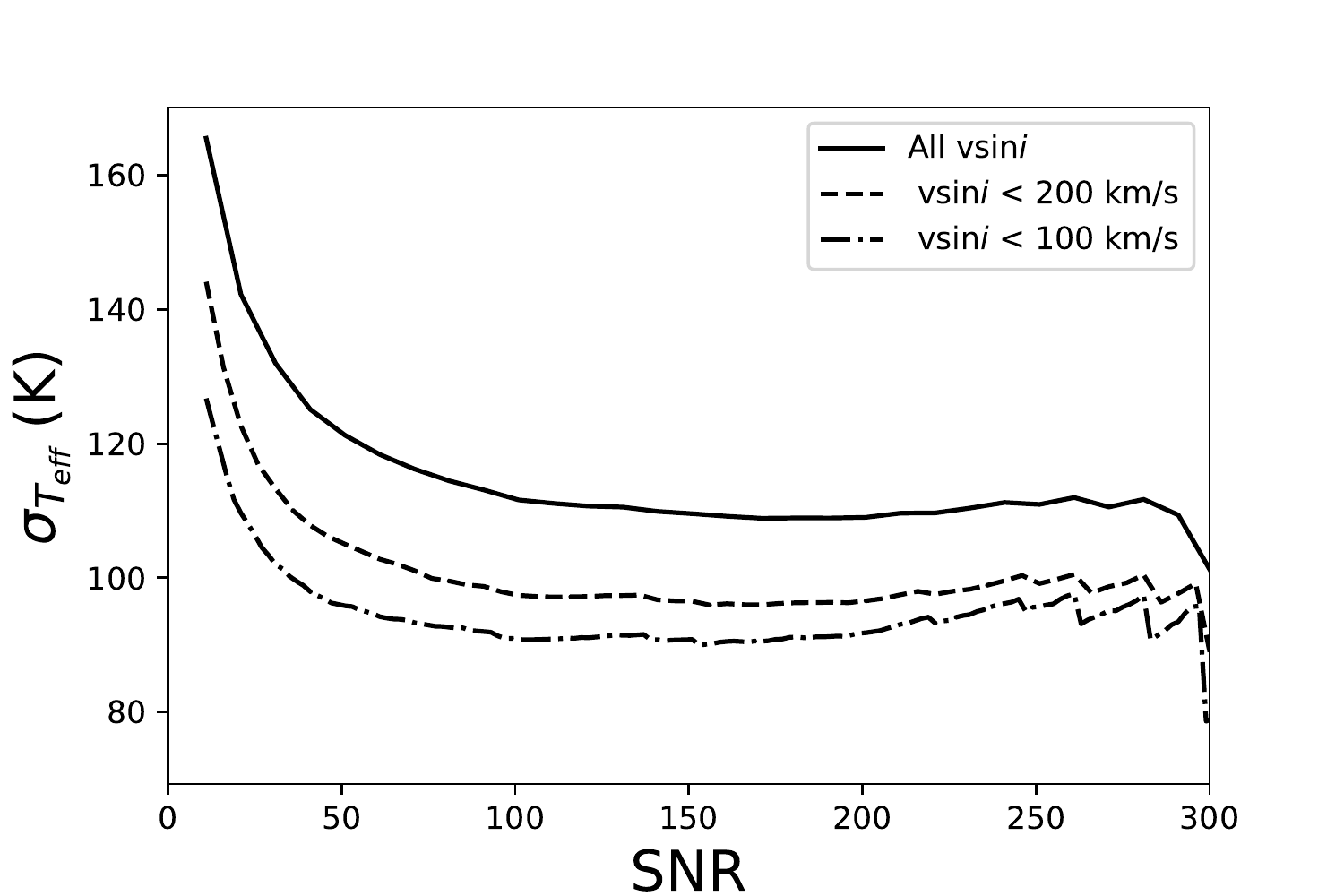}
		\includegraphics[scale=0.45]{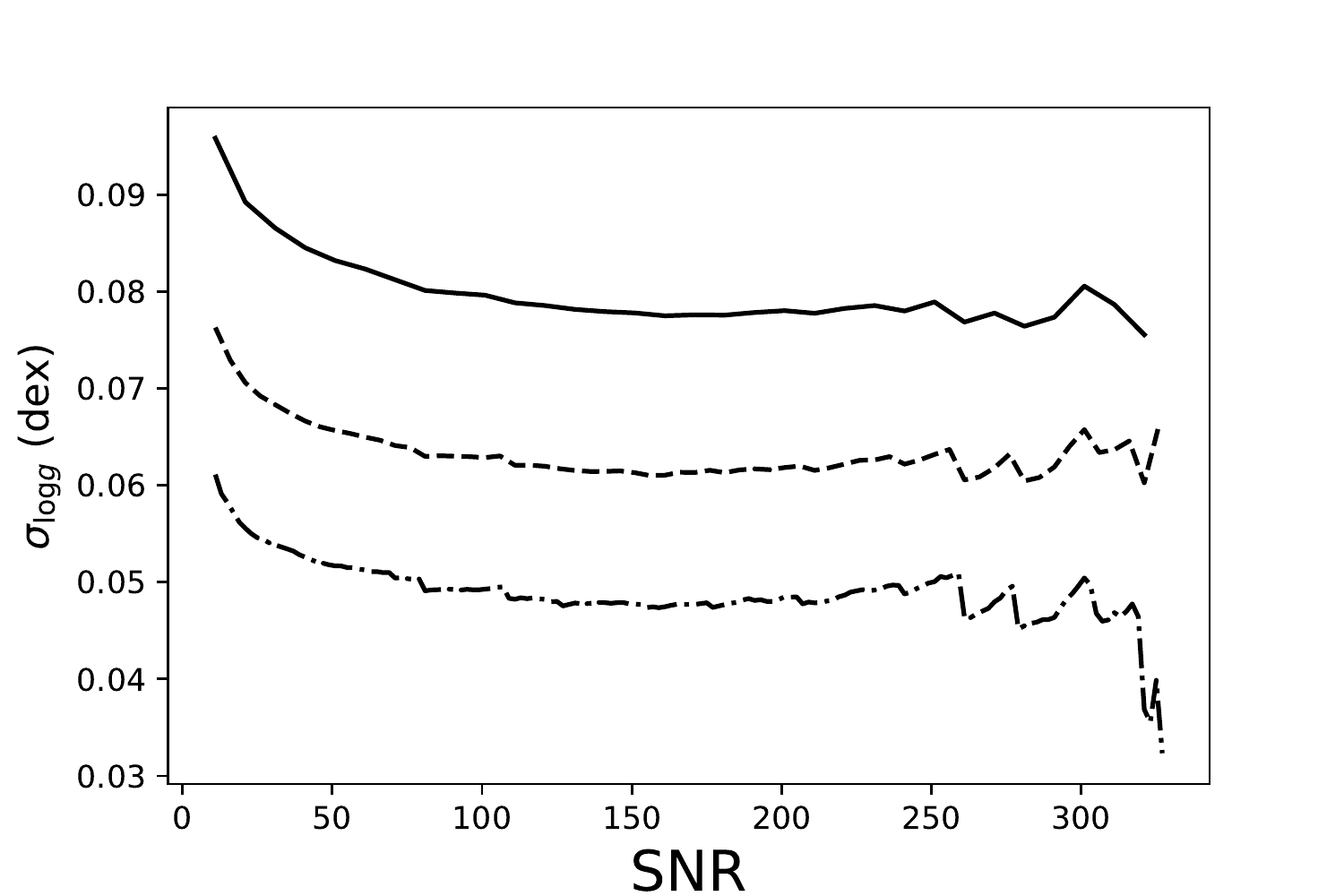}\\
		\includegraphics[scale=0.45]{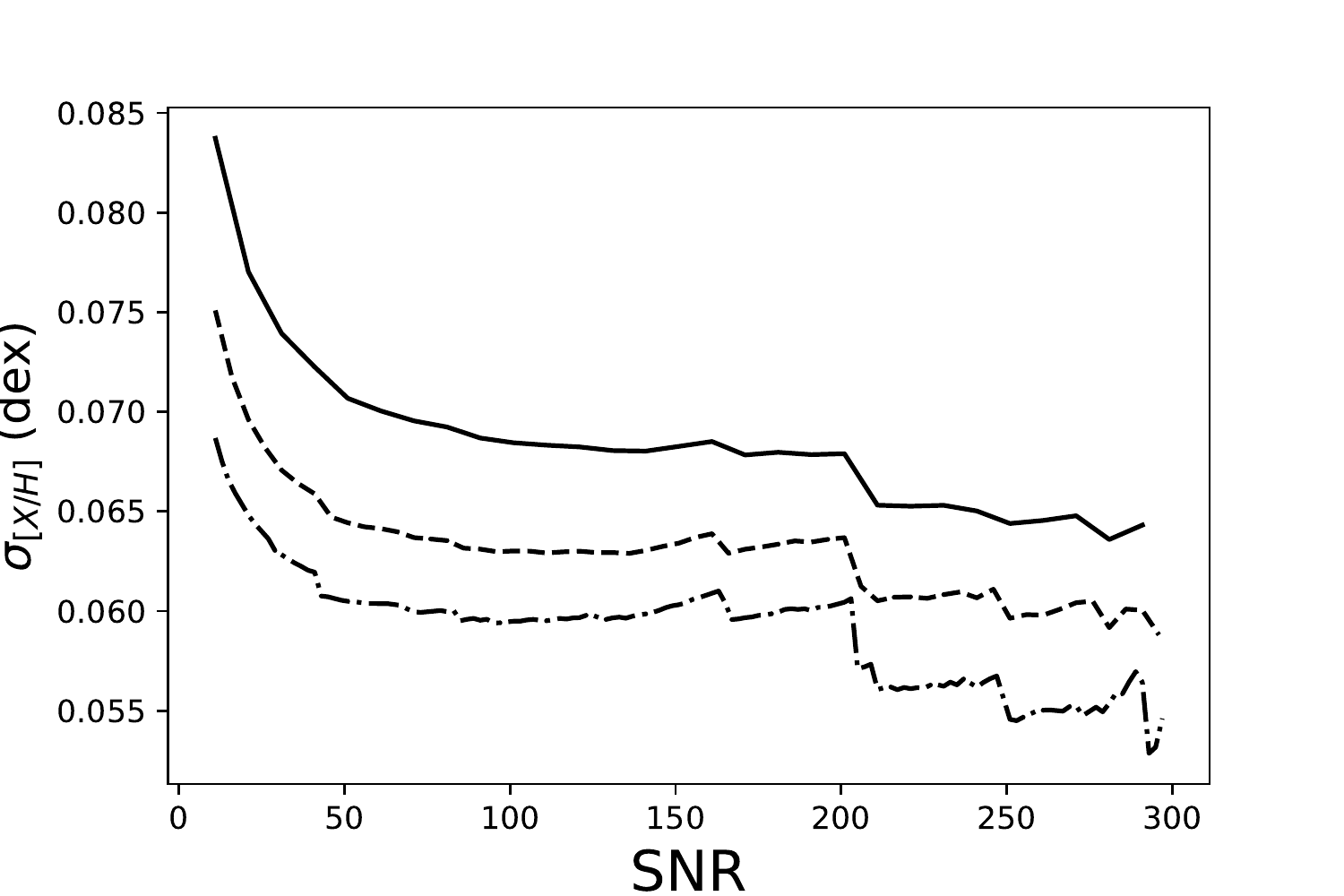}
		\includegraphics[scale=0.45]{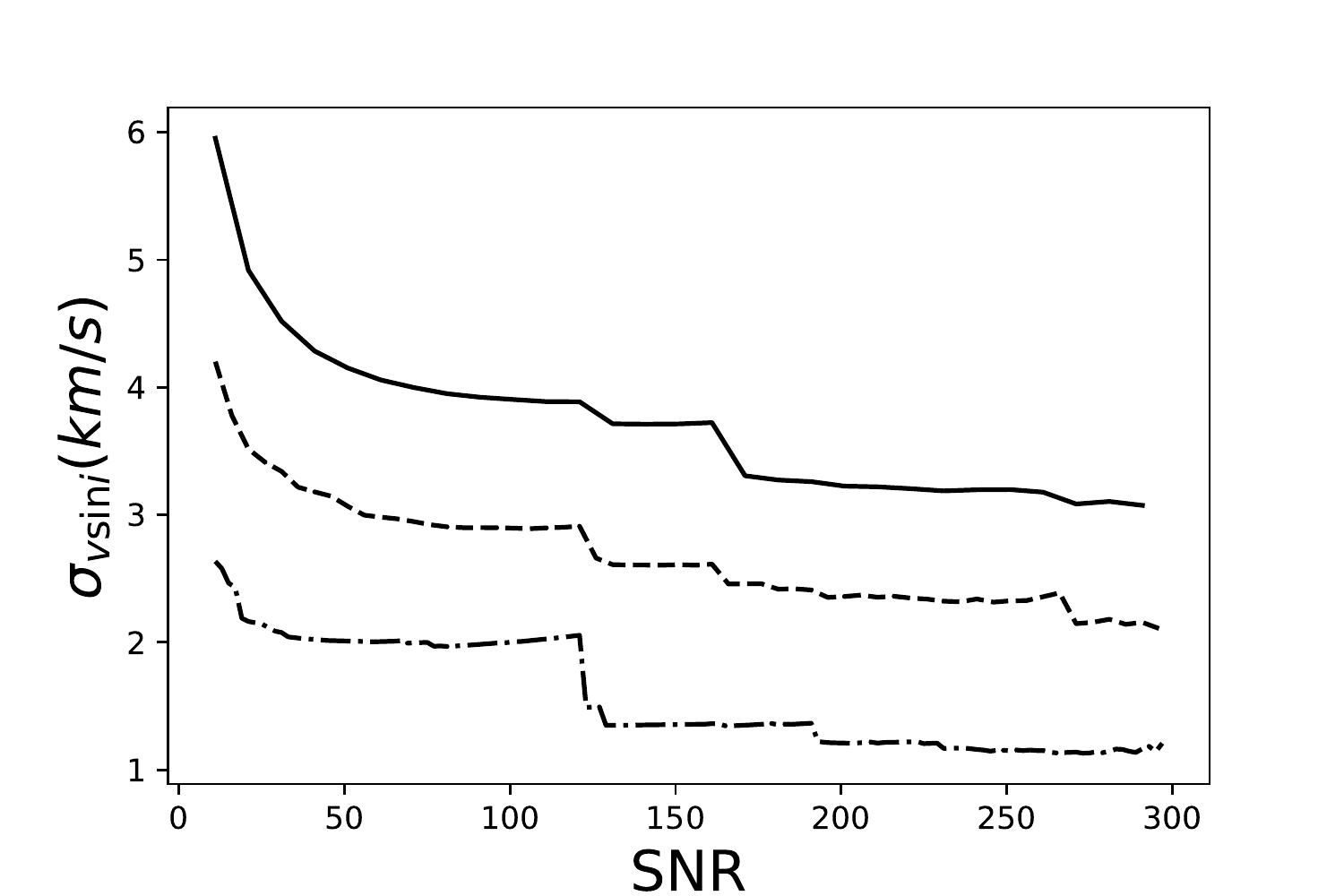}
		\caption{Average error bars for the observation predicted stellar parameters as a function of the signal to noise ratio and for different ranges of stellar rotation. }
		\label{fig:SNR}
	\end{figure*}

	In order to verify the effect of the noise on the predicted parameters, Fig.~\ref{fig:SNR} displays the variation of the accuracy of the predicted values with respect to the input SNR. The figure also displays the observations depending on the values of \vsini. The reason for that is that increasing \vsini\ induces blending in the spectra and thus less information to be used in the prediction. This is reflected in the case of low \vsini\ for which the predicted values are found to be more accurate than the case of large \vsini.

	\section{Extrapolating to other spectral types}
	\label{fgk}
	In order to verify how universal the results are, we checked that the optimization of the code is not dependent on wavelength and/or spectral type, we also tested the procedure on FGK stars. To do that, we have calculated a TDB specific for FGK stars using the parameters displayed in Tab.~\ref{FGK-limits}. The wavelength range was selected to coincide with the one of \cite{S4n}. This range is sensitive to all the concerned stellar parameters. 
	\begin{table}
		\centering
		
		\begin{tabular}{|c|c|} 
			\hline
			Parameters & Range \\
			\hline
			\Teff\ (K) & [4\,000,7\,000]  \\
			\logg\ (dex)  & $[3.0, 5.0]$  \\
			\met   (dex)  & $[-1.5 , 1.5]$  \\ 
			\vsini\ (\kms) & $[0, 100] $  \\
			$\lambda$ ($\AA$) &  [5\,000, 5\,400]    \\
			$\lambda/\Delta \lambda$ & 60\,000  \\
			\hline
		\end{tabular}
		\caption{Ranges of the parameters used for the calculation of the FGK synthetic spectra TDB.}
		\label{FGK-limits}
	\end{table}
	
	A database of 50\,000 random synthetic spectra with known stellar labels is used in the training. About 20\,000 Test data, with and without noise, were calculated in the same range of Tab.~\ref{FGK-limits} to be used for verification. The optimal NNs that were introduced in Sec.~\ref{results} were used again, as a proof of concept, for the FGK TDB. The results are displayed in Tab.~\ref{FGK-results} for Training, Validation, and tests.

	\begin{table*}
		\centering
		\begin{tabular}{|l|c|c|c|c|} 
			\hline
			& Training& Validation & Test (no noise)& Test (noise)\\
			\hline
			$\sigma_{\Teff}$ (K)& 59 & 62& 62 & 82\\
			$\sigma_{\logg}$ (dex)& 0.04 & 0.05 & 0.05 & 0.07 \\
			$\sigma_{\vsini}$ (\kms)&0.40&0.50&0.55&0.90  \\
			$\sigma_{\met}$ (dex)& 0.04&0.05&0.05&0.06 \\
			\hline
		\end{tabular}
		\caption{Derived standard deviation for each parameters using the TDB for FGK stars. }
		\label{FGK-results}
	\end{table*}
	
	Because of the low rotational velocities for FGK stars (\vsini $<$ 100 \kms), the results are more accurate. That is not surprising because \vsini\ drastically affects the shape of the lines as in A stars. The derived errors on the stellar parameters are found to be 82 K, 0.07 dex, 0.90 \kms and 0.06 dex for \Teff, \logg, \vsini and \met, respectively (Tab.~\ref{FGK-results}). These results are very promising, but we should be aware of the complications that would arise when using real observations, especially in the case of the cool M stars. These stars have been analyzed in the context of exoplanets search (\cite{CARMENES2,CARMENES1}) and show complications in their spectra mainly related to the continuum normalization. Adapting the data preparation and the CNN will be inevitable in order to take into account these effects. These results also show that when deriving the stellar parameters for specific spectral types, the wavelength region should be selected according to these spectral lines/bands most sensitive to the variations of the parameters one seeks.

	\section{Discussion and Conclusion}
	\label{disc}

	The purpose of this work is not only to find the best tool for the accurate prediction of parameters but also to show the steps that should be taken in order to reach the optimal selection of the CNN parameters. Often scientists use DL as a black box without explaining the choice of the parameters and/or architecture. In this manuscript, we have explained the reason for selecting specific hyperparameters while emphasizing the pedagogical approach.  To have a more effective tool, one should change the architecture of the model. The architecture of the model depends on the type and range of the input. In this work we have fixed the architecture and iterated on the hyperparameters only. 
	
	Sections \ref{teff} to \ref{vsini} show that for each stellar parameter, the setup of the network should be changed. This means that for a specific network and a specific stellar parameter, a study should be made to find the optimal configuration of hyperparameters. This is due to the contribution of the specific stellar parameter on the shape of the input spectrum. Using the PCA decomposition, we have reduced the size of the input parameters to only 50 points per spectrum while keeping more than 99.5\% of the information. This is recommended in case of large databases and wide wavelength range and could avoid the use of extra pooling layers in the network. This projection technique is not only applicable for AFGK stars but can also be used for cooler stars. Further, (\cite{2016ApJ...822...97H,dms}), \cite{2018MNRAS.476.1120S} have applied a projection pursuit regression model based on the independent component analysis compression coefficients to derive \Teff, \logg, and \met\ of M-type stars.
	
	Although the CNN architecture was not optimized, we were able, using a strategy of finding the best hyperparameters, to reach a level of accuracy that is comparable to other adopted techniques. In fact, we found for A stars, an average accuracy of 0.08 dex for \logg, 0.07 dex for \met, 3.90 \kms\ for \vsini, and 127 K for \Teff. In the case of stars with \vsini\ less than 100 \kms, we found the accuracy to be 90 K, 0.06 dex, 0.06 dex and 2.0 \kms, for \Teff, \logg, \met\ and \vsini, respectively. These accuracy values are signal to noise dependant and reduce as long as the signal to noise increases. Extrapolating the technique to FGK stars also shows that the same network could be applied to different spectral types and different wavelength ranges.

	The technique that we followed in this paper could be transferable to any classification problem that involves neural network. 
	In the future we plan to develop a strategy to find the best CNN architecture depending on the input data and the type of the predicted parameters. Once the architecture and the configuration of the parameters is settled, we will be testing the procedure on observational spectra as we did in \cite{S4n}, \cite{dms}, \cite{Gebran} and \cite{2019OAst...28...68K}. Using only observational data or a combination of synthetic spectra and real observations with well known parameters will allow us to constrain the derived stellar labels while minimizing the critical synthetic gap (\cite{Fabro}).  One more criterion that should be taken into account is when applying this technique to real observations, thorough data preparation work should be done to take into account the characteristics of each spectral type (e.g., continuum normalization in M and giant stars, and low number of lines in hot stars.).

{\small
	\bibliographystyle{ieee_fullname}
	\bibliography{main}
}

\end{document}